\documentclass[fleqn,usenatbib]{mnras}

\usepackage[T1]{fontenc}

\DeclareRobustCommand{\VAN}[3]{#2}
\let\VANthebibliography\thebibliography
\def\thebibliography{\DeclareRobustCommand{\VAN}[3]{##3}\VANthebibliography}

\usepackage{graphicx}	
\usepackage{amsmath}	
\usepackage{amssymb}	

\usepackage{amssymb}
\usepackage{longtable,ltxtable,booktabs}

\usepackage{multirow}
\usepackage{color}

\def\approxge{\mathrel{\raise1.16pt\hbox{$>$}\kern-7.0pt \lower3.06pt\hbox{{$\scriptstyle\approx$}}}} 
\def\approxle{\mathrel{\raise1.16pt\hbox{$<$}\kern-7.0pt \lower3.06pt\hbox{{$\scriptstyle \approx$}}}} 

\title[Orbital Evolution of Contact NS-WD binaries]{Orbital Evolution of Neutron-Star -- White-Dwarf Binaries by Roche-Lobe Overflow and Gravitational Wave Radiation}

\author[]{Shenghua Yu$^{1}$\thanks{shenghuayu@bao.ac.cn}, Youjun Lu$^{2,3}$\thanks{luyj@nao.cas.cn}, and C. Simon Jeffery$^{4}$\thanks{Simon.Jeffery@armagh.ac.uk}\\
$^1$CAS Key Laboratory of FAST, National Astronomical Observatories, Chinese Academy of Sciences, 20A Datun Road, \\ Beijing 100101, China \\
$^2$CAS Key Laboratory for Computational Astrophysics, National Astronomical Observatories, Chinese Academy of Sciences, \\ 20A Datun Road, Beijing 100101, China \\
$^3$School of Astronomy and Space Sciences, University of Chinese Academy of Sciences, 19A Yuquan Road, Beijing 100049, China \\
$^4$Armagh Observatory and Planetarium, College Hill, Armagh BT61 9DG, N. Ireland
}

\date{Accepted XXX. Received YYY; in original form ZZZ}

\pubyear{2020}

\begin{document}
\label{firstpage}
\pagerange{\pageref{firstpage}--\pageref{lastpage}}
\maketitle

\begin{abstract}
We investigate the effects of mass transfer and gravitational wave (GW) radiation
on the orbital evolution of contact neutron-star-white-dwarf (NS-WD) binaries, and the detectability
of these binaries by space GW detectors (e.g., Laser Interferometer Space Antenna, LISA; Taiji;
Tianqin). A NS-WD binary becomes contact when the WD component fills its Roche lobe, at which
the GW frequency ranges from $\sim0.0023$ to $0.72$\,Hz for WD with masses $\sim0.05-1.4{\rm M}_{\odot}$.
We find that some high-mass NS-WD binaries may undergo direct coalescence after unstable mass
transfer. However, the majority of NS-WD binaries can avoid direct coalescence because mass transfer after contact can lead to a reversal of the orbital evolution.
Our model can well interpret the orbital evolution of the ultra-compact X-ray source   4U\,1820--30.
For a $4$-year observation of 4U\,1820--30, the expected signal-to-noise-ratio (SNR) in GW characteristic strain is $\sim11.0/10.4/2.2$ (LISA/Taiji/Tianqin).
The evolution of GW frequencies of NS-WD binaries depends on the WD masses.
NS-WD binaries with masses larger than 4U\,1820--30 are expected to be detected with significantly larger SNRs.
For a $(1.4+0.5){\rm M}_{\odot}$ NS-WD binary close to contact, the expected SNR for a one week observation is $\sim27/40/28$ (LISA/Taiji/Tianqin).
For NS-WD binaries with masses of $(1.4+\gtrsim1.1){\rm M}_{\odot}$, the significant change of GW frequencies and amplitudes can be measured, and thus it is possible to determine the binary evolution stage.
At distances up to the edge of the Galaxy ($\sim100$\,kpc), high-mass NS-WD binaries will be still detectable with SNR$\gtrsim$1.
\end{abstract}


\section{Introduction}
\label{intro}

Ultracompact X-ray binary stars (UCXB) are believed to be neutron-star-white-dwarf (NS-WD) binaries with significant mass transfer.
They are not only excellent laboratories for testing binary-star evolution but are also important gravitational wave (GW) sources.
Studies of their orbital parameters (mass, orbital period and eccentricity), chemical composition and spatial distribution may provide important information and clues to understand both the accretion processes of compact binaries under extreme conditions and the formation and evolution of double compact binaries, including various physical processes involved in, for example, common envelope evolution.
Some NS-WD binaries have extremely short orbital periods (10 - 80 minutes) and may radiate GW in the frequency band of $10^{-4}-1$\,Hz, and will be interesting sources for the Laser Interferometer Space Antenna (LISA) \citep{Nelemans09, Nelemans10}, Taiji \citep{Ruan20}, and Tianqin \citep{Wang19}.

There are  $\sim13$  UCXBs with known orbital periods ($5$ in globular clusters), among which 4U\,1820--30 has the shortest orbital period ($P_{\rm orb}$), coincident with  regular coherent X-ray bursts of $685$\,s \citep{Stella87, Verbunt87}.
4U\,1820--30 is located in the globular cluster NGC 6624 near the Galactic bulge \citep{Grindlay76, Stella87} at a distance of $7.6\pm0.4$\,kpc \citep[][]{Kuulkers03} or $8.4 \pm 0.6$\,kpc \citep[][]{Valenti07} from the Earth.
Radio observations of 4U\,1820--30 indicate that there is one single pulsar with rotation period $=5.4$\,ms near the position of the X-ray source \citep{Geldzahler83, Grindlay86, Biggs94}. In addition, it is suggested that the strong UV and visible flux between 140--430 nm observed by HST from 4U\,1820--30 is most likely coming from the heated accretion disk or heated side of a WD \citep{King93}.
Based on the X-ray spectrum analysis using the multiple thermonuclear explosion model, the mass and radius of the NS component are estimated to be $1.58 \pm 0.06 {\rm M}_{\odot}$ and $9.1 \pm 0.4$\,km \citep{Guver10}, respectively. However, these estimates may be model dependent. The companion is commonly accepted to be a helium WD with mass in the range $0.06-0.07 {\rm M}_{\odot}$ \citep{Rappaport87,Suleimanov17}, according to the Roche lobe filling WD model.

The orbital period of 4U\,1820--30 is evolving with time as indicated by X-ray observations.
The change is {\it measured} as $\dot{P}_{\rm orb}/P_{\rm orb}=(-0.53\pm 0.03)$ $\times 10^{-7}$ yr$^{-1}$ (\citealt[][]{Peuten14}; see also estimates by \citealt{vanderKlis93}a; \citealt{vanderKlis93b}b; \citealt{Tan91, Morgan88, Chou01}), where $\dot{P}_{\rm orb}$ is the time derivative of $P_{\rm orb}$,
which means the orbit is shrinking.
These measurements contradict an original {\it prediction} by \citet{Rappaport87} that $\dot{P}_{\rm orb}/P_{\rm orb}\approx (1.1\pm0.5)$ $\times 10^{-7}$ yr$^{-1}$,
implying that the orbital separation is increasing as a consequence of mass transfer.
Various mechanisms  have been proposed  to explain the orbital evolution of 4U\,1820--30, including the binary acceleration in the gravity potential of globular cluster (\citealt{vanderKlis93}a, \citealt{Chou01}), the influence of vertical structure of mass flow at the edge of accretion disk (\citealt{vanderKlis93b}b), the influence of a third companion or the mass donating star being a burning He star (\citealt{vanderKlis93}a).
The evolution status of 4U\,1820--30 remains unclear partly because of the contradictory observational results.

The orbital evolution of UCXBs such as 4U\,1820--30
is in general complicated by the participation of a Roche lobe-filling WD in the orbital angular momentum evolution, as well
as other physical processes including tidal effects, magnetic fields, and spin-orbit coupling.
Space-based GW observations can provide a unique way to study the orbital evolution of NS-WD binaries since in the Milky Way such binaries, with typical orbital periods of a few to hundreds of seconds, radiate GWs.
It is useful to identify some NS-WD binaries as verification sources for space-based GW detectors.
GW observations of UCXBs will be important for the investigation of the coalescence of individual sources \citep{Tauris18} and in searching for additional short-period  UCXBs.

In this paper, we develop a  code (\S~\ref{sec_method}) to simulate the evolution of NS-WD binaries where Roche-lobe overflow occurs.
We explore the evolution of 4U\,1820--30 and other binaries with different masses in \S~\ref{sec_results}.
We also investigate the detectability of such NS-WD binaries by space GW detectors, such as LISA, Taiji, and Tianqin (\S~\ref{sec_results}.7).
Some discussion and conclusions are given in \S~\ref{sec_discussion} and \ref{sec_conclusion}.

\section{Method}
\label{sec_method}

\subsection{Orbital evolution of a mass-transferring binary}

The orbital angular momentum evolution ($\dot{J}_{\rm orb}$) of a NS+WD binary is mainly controlled by the changes due to GW radiation $\dot{J}_{\rm GR}$, mass transfer $\dot{J}_{\rm mt}$,
and the coupling between the spin of the (primary) accretor and the binary orbit angular momentum $\dot{J}_{\rm so}$, i.e.,
\begin{equation}
\dot{J}_{\rm orb}=\dot{J}_{\rm GR}+\dot{J}_{\rm mt}+\dot{J}_{\rm so}.
\label{eq_jorb1}
\end{equation}
The orbital separation of an initially detached NS-WD binary shrinks due to the GW radiation.
The change of orbital angular momentum due to GW radiation is
\begin{equation}
\frac{\dot{J}_{\rm GR}}{J_{\rm orb}}=-\frac{32}{5}\frac{G^{3}}{c^{5}}\frac{m_{1}m_{2}M}{a^{4}}
\label{eq_jgw}
\end{equation}
\citep{Maggiore08}, where $G$ is the gravitational constant, $c$ the speed of light, $m_{1}$ and $m_2$ are the masses of the primary (NS) and secondary (WD) components, respectively, $M=m_{1}+m_{2}$ the total mass, and $a$ the semimajor axis. The total orbital angular momentum is
\begin{equation}
J_{\rm orb}=m_{1}m_{2}\sqrt{Ga/M},
\label{eq_jorbdef}
\end{equation}
assuming a circular orbit.

With orbital decay due to GW radiation, the  Roche lobe will eventually shrink to reach the material surface of the (secondary) WD, which will then transfer to the vicinity of the primary NS via the first Lagrangian point.
It will form an accretion disk around the NS if the NS radius $r_{1}$ is smaller than the minimum distance $r_{\rm min}$ between the closest approach of the mass flow and the NS center.
\citet{Lubow75} estimated $r_{\rm min}$ by detailed analysis of the gas dynamics for semi-detached binary systems. For simplicity, we adopt the fitting formula given by \citet{Nelemans01a} to estimate $r_{\rm min}$.

The angular momentum transferred with the mass flow from the first Lagrangian point to the neighbourhood of the accreting NS can cause a change of the orbital angular momentum $\dot{J}_{\rm mt}$, which can be written as
\begin{equation}
\dot{J}_{\rm mt}=\dot{m}_{1}\sqrt{Gm_{1}R_{\rm h}},
\label{eq_jmf}
\end{equation}
where $R_{\rm h}$ is an equivalent radius defined for the matter orbiting the accreting NS. If a fully developed accretion disk is present, the angular momentum caused by the mass flow is cancelled by a reverse flow \citep{Lubow75}. In such a case, we set $R_{\rm h}=0$.
Otherwise, we employ $R_{\rm h} = a r_{\rm h}$, where $r_{\rm h}$ is a parameter, depending on the mass ratio.
By fitting the results of numerical computations, it can be expressed as $r_{\rm h} \approx a \left[0.0883+0.04858 \log q + 0.11489 \log^{2} q - 0.020475 \log^{3} q \right]$, with $0<q=m_{2}/m_{1}\leqslant 1$ being the mass ratio \citep{Verbunt88}.

The coupling between the spin of the primary NS (accretor) and the binary orbit is probably caused by a torque from dissipative coupling, tidal, or magnetic. We parameterize this torque in terms of the synchronization timescale $\tau_{\rm s}$ of the accretor \citep{Marsh04}, i.e.,
\begin{equation}
\dot{J}_{\rm so}=\frac{km_{1}r_{1}^{2}}{\tau_{\rm s}}(\omega_{\rm s}-\omega_{\rm o}),
\label{eq_jso}
\end{equation}
which is proportional to the difference between the accretor's rotation frequency $\omega_{\rm s}$ and the orbital angular frequency $\omega_{\rm o}$.
In Eq.\,\ref{eq_jso},  the term $km_{1}r_{1}^{2}$ is the moment of inertia of the accretor  and $r_{1}$ is the radius of the accretor (neutron star).
$k$ is a function of the density profile and takes values between 0.2 and 0.4 for polytropes with index $n$ between 1.5 and 0.
We here adopt $k=0.2$ ($n\approx1.5$) for simplicity.
Since neutron stars are better represented by $n\approx0.5 - 1$, this approximation should be investigated in future work.

The accretor's rotation frequency evolves with the orbital angular frequency as
\begin{equation}
%
%
\dot{\omega}_{\rm s} =\dot{\omega}_{\rm o}+\dot{\Omega}_{\rm so} 
=\lambda \omega_{\rm s}\frac{\dot{m}_{2}}{m_{1}}-\frac{\dot{J}_{\rm orb}}{k m_{1}r_{1}^{2}}-\frac{\Omega_{\rm so}}{\tau_{\rm s}}
%
%
\label{eq_omegas}
\end{equation}
where $\lambda=1+2\frac{{\rm d ln} r_{1}}{{\rm d ln} m_{1}}+\frac{{\rm d ln} k}{{\rm d ln} m_{1}}$, $\Omega_{\rm so} = \omega_{\rm s} -\omega_{\rm o}$.
The tidal timescale $\tau_{\rm s}$ depends on the most effective dissipative mechanisms available \citep[cf.][]{Zahn77,Eggleton98,Preece19}.
Since it was explicity designed for double-degenerate binary stars, we adopt the expression of \citet{Campbell84}
\begin{equation}
\tau_{\rm s,tid}=1.3 \times 10^{7}q^{2}\Big(\frac{a}{r_{2}}\Big)^{6}\Big(\frac{m_{2}/{\rm M_{\odot}}}{L_{2}/{\rm L_{\odot}}}\Big)^{5/7} ~~{\rm yr},
\label{eq_stid}
\end{equation}
where $L_{2}$ is the luminosity of secondary.
Another important factor to affect the synchronization timescale is the dissipation of  electrical currents within the donor star induced by the accretor's magnetic field \citep{Campbell83}, which can be expressed as
\begin{equation}
\tau_{\rm s,mag}=2 \times 10^{6}~m_{1} r_{1}^{-4}r_{2}^{-5}a^{6}B^{-2} ~~{\rm yr},
\label{eq_synch1}
\end{equation}
where $r_{2}$ is the radius of mass donor (WD), and $B$ is the surface magnetic field of the accretor in Gauss.
All other quantities in Eqs.\,\ref{eq_stid} and \ref{eq_synch1} are in solar units.
The timescale $\tau_{\rm s}$ may be very short for magnetized NSs. Considering a semidetached binary just at the onset of mass transfer,
for example, with $m_{1}=1.4 {\rm M}_{\odot}$,  $r_{1}=1.44\times 10^{-5} {\rm R}_{\odot}$,  $m_{2}=0.2{\rm M}_ {\odot}$, $r_{2}=0.020954 {\rm R}_{\odot}$,
semimajor axis of $a=0.914 {\rm R}_{\odot}$, and $B=10^{12}$\,G, we have $\tau_{\rm s} \simeq \tau_{\rm s,mag}\approx 9500$ yr.
For the same binary, if employing the luminosity $L_{2}=0.1 L_{\odot}\approx 3.845\times10^{32}$ ergs s$^{-1}$
according to the optical identification of He WD-pulsar binaries in the globular cluster 47 Tucanae \citep{Cadelano15},
we have $\tau_{\rm s} \simeq \tau_{\rm s,tid}\approx 3.0\times10^{9}$ yr. However, if taking into account the luminosity
caused by the disc and outflow mass, the luminosity $L_{2}$ may be $L_{2}=10^{7}$ $L_{\odot}$ \citep{Metzger12,Margalit16},
leading to a much smaller synchronization time $\tau_{\rm s,tid}\approx 5800$ yr and strong coupling.
We here take the secondary luminosity $L_{2} < 1L_{\odot}$, neutron star rotation frequency $\sim 0.1-1000$ Hz  and magnetic field  $\sim 10^{8}-10^{12}$ G  (http://www.atnf.csiro.au/research/pulsar/psrcat, \citet{Manchester05}).

The accretor may be spun up to its breakup rate at which point spin-orbit coupling will presumably
strengthen markedly through bar-mode type instabilities or the shedding of mass. We model this by following \citet{Marsh04}. The breakup angular frequency is approximately equal to the Keplerian orbital angular frequency at the surface of the accretor
\begin{equation}
\omega_{\rm k}=\sqrt{\frac{Gm_{1}}{r_{1}^{3}}}.
\end{equation}
The time derivative of $\omega_{\rm k}$ is
\begin{equation}
\dot{\omega}_{\rm k}=\frac{\omega_{\rm k}}{2}(1-3\varsigma_{1})\frac{\dot{m}_{1}}{m_{1}},
\end{equation}
with $\varsigma_{1}=\frac{{\rm d ln}r_{1}}{{\rm d ln}m_{1}}$.
In this case, we force $\dot{\omega}_{\rm s}=\dot{\omega}_{\rm k}$.

On the other hand, the time derivative of orbital angular momentum $\dot{J}_{\rm orb}$ can be obtained by its definition, i.e.,
\begin{equation}
\frac{\dot{J}_{\rm orb}}{J_{\rm orb}}=\frac{\dot{a}}{2a}+\frac{\dot{m}_{1}}{m_{1}}
+\frac{\dot{m}_{2}}{m_{2}}-\frac{\dot{M}}{2M}.
\label{eq_jorb2}
\end{equation}
During the mass transfer phase, the mass donor $m_{2}$ loses an amount of mass $-{\rm d}m_{2}$ in a unit time interval. A fraction $\alpha {\rm d}m_{\rm 2}$ ($0 \leqslant \alpha \leqslant 1$) is accreted
by the companion, implying $\dot{m}_{1}=\alpha \dot{m}_{2}$. The remainder ($(1-\alpha) {\rm d}m_{\rm 2}$ is assumed to be ejected from the binary system; thus $\dot{M}=(1-\alpha)\dot{m}_{2}$. If $\alpha=1$, the total mass of the system is conserved, and if $\alpha <1$, the binary gradually loses mass. In addition, we assume that the mass ejected from the binary has a specific angular momentum $j\equiv {\rm d}J/{\rm d}M=\beta J/M$ \citep{Pols94}. For simplicity, we set $\beta=1$, thus ${\rm d}J/J={\rm d}M/M$ and $\dot{J}_{\rm orb}/J_{\rm orb}=\dot{M}/M$.

We obtain the evolution of the semimajor axis of the NS-WD binary by combining Eqs.~\eqref{eq_jorb1} and \eqref{eq_jorb2} as
\begin{equation}
\begin{split}
\frac{\dot{a}}{2a}=& -\frac{32}{5}\frac{G^{3}}{c^{5}}\frac{m_{1}m_{2}M}{a^{4}} 
%
+\left[1 - \alpha q
- \left(1-\alpha\right) \frac{q}{1+q} \left(\beta+\frac{1}{2}\right)  \right. \\
&  \left.- \alpha \sqrt{(1+q)r_{\rm h}}  \right ]\frac{-\dot{m}_{2}}{m_{2}} 
%
+\frac{km_{1}r_{1}^{2}}{\tau_{\rm s}J_{\rm orb}}(\omega_{\rm s}-\omega_{\rm o}).
\label{eq_adot}
\end{split}
\end{equation}
Noting that $-\dot{m}_2$ is positive,
the binary orbit always decays due to GW radiation, but may expand  due to the last two terms.
Eq.~\eqref{eq_adot} differs from \citet[][Eq. 6]{Marsh04}, which was used to calculate the orbit evolution of double WDs by assuming mass conservation. Eq.~\eqref{eq_adot} includes mass and angular momentum loss from the binary.
and reduces to \citet[][Eq. 6]{Marsh04} if $\alpha=1$.

In contrast to previous studies (e.g., \citet{Verbunt88,Marsh04}), we also consider the evolution of mass transfer by considering the structure evolution of the WD. Assuming that the self-gravity of a WD is balanced by the electron degenerate pressure $P_{\rm e}$, the structure of a WD along its radius $r_{2}$ can be numerically obtained by solving the hydrostatic equation and Poisson's Eq., i.e.,
\begin{equation}
\frac{{\rm d} P_{\rm e}}{{\rm d} r_{2}}=-\frac{{\rm d} \Phi}{{\rm d} r_{2}}\rho_{2},
\label{eq_hydrostatic}
\end{equation}
\begin{equation}
\nabla^{2} \Phi= 4\pi G \rho_{2},
\label{eq_poisson}
\end{equation}
where $r_{2}$ is the radius of a WD and $\rho_{2}$ is its mass density. The degenerate pressure of an electron fermion gas $P_{\rm e}$ and mass density of a WD can be written as \citep{Kippenhahn90}
\begin{equation}
P_{\rm e}=\frac{\pi m_{\rm e}^{4}c^{5}}{3h^{3}}\cdot \left[(2x^{3}-3x)\sqrt{x^{2}+1}+3 \ln (x+\sqrt{x^{2}+1})\right],
\label{eq_gpe}
\end{equation}
and
\begin{equation}
\rho_{2}=\mu_{\rm e}m_{\rm u} \frac{8\pi m_{\rm e}^{3}c^{3}}{3 h^{3}}x^{3}.
\label{eq_denm}
\end{equation}
where $m_{\rm e}$ is the rest mass of  electron, $m_{\rm u} \approx 1.66\times10^{-24}$\,g the atomic mass unit, $h$ the Planck constant, $x=p_{\rm m}/m_{\rm e}c$ with $p_{\rm m}$ being a certain momentum. All phase cells below the momentum $p_{\rm m}$ are occupied by two electrons, and all phase cells above $p_{\rm m}$ are empty. For a pure electron gas, the mean weight $\mu_{\rm e}$ is
\begin{equation}
\mu_{\rm e} \approx \left(X+\frac{1}{2}Y+\frac{1}{2}Z\right)^{-1},
\label{eq_meanwe}
\end{equation}
where $X$, $Y$, and $Z$ represent the fraction of the mass density contributed by hydrogen, helium, and heavier elements in a star, respectively. We have $X+Y+Z=1$.

\begin{figure}
\hspace*{-0.8cm}
\centering
\includegraphics[width=10cm,clip,angle=0]{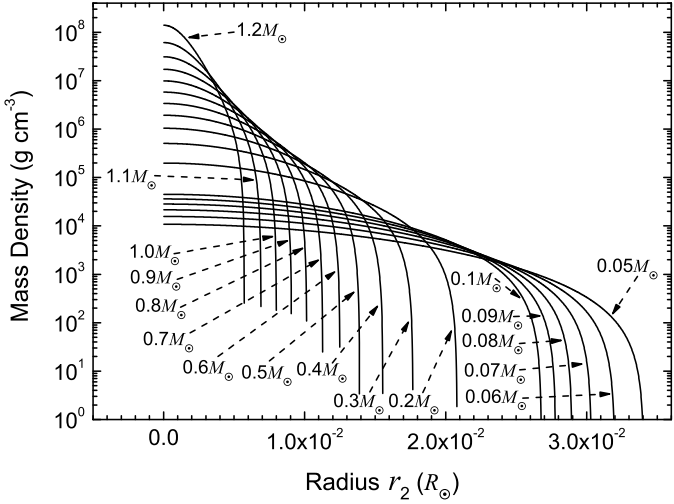}
\caption{The mass density as a function of radius of white dwarfs in a computation
grid of white dwarf masses.
}
 \label{fig1}
\end{figure}

Significant mass transfer takes place when the surface of a WD reaches its Roche lobe. The WD then will expand as a result of losing mass. The condition for the onset of mass transfer can be written as $r_{2}=r_{\rm 2L}$ where $r_{\rm 2L}$ is the Roche lobe radius of the mass donor. We estimate $r_{\rm 2L}$ according to the following approximation \citep{Eggleton83}
\begin{equation}
r_{\rm 2L}=a\frac{0.49q^{2/3}}{0.6q^{2/3}+\rm
ln(1+\it q^{\rm 1/3} )},
\label{eq_rl}
\end{equation}
where $q$ ($= m_{2}/m_{1}$) is the mass ratio.
The mass fraction which flows outside of the Roche lobe per unit time, i.e., the mass transfer rate and mass loss from the binary system, can be roughly estimated as the product of volume ($\approx r_{2}^{3}-r_{\rm 2L}^{3}$) of overflow matter outside of the Roche lobe and mass density $\rho_{2}$ of the overflow matter, i.e.,
\begin{equation}
\dot{m}_{2}\approx \frac{\Delta m_{2}}{\Delta t}=-\frac{4\pi}{3}\frac{\rho_{2}(r_{2}^{3}-r_{\rm 2L}^{3})}{\Delta t},
\label{eq_mtrate}
\end{equation}
where $\Delta t$ is a small time interval (time step in the simulations).
The change of $r_{2}$ and $\rho_{2}$ with respect to time are obtained by
numerically solving Eqs.~\eqref{eq_hydrostatic}$-$\eqref{eq_meanwe}.
Figure~\ref{fig1} shows mass density structure as a function of radius $r_{2}$ of WDs with different masses.

In order to determine the fraction of matter which is accreted by
the NS, we assume that (1) if the accretion luminosity
is less than the Eddington limit $L_{\rm Edd}$, all matter lost from the WD can be accreted by the NS;
(2) if the accretion luminosity is larger than the Eddington limit
and the diffusion timescale of outward photons released in the accretion
flow is slower than the timescale of photons swept inward by the accretion
flow itself, the excess fraction of energy is absorbed in the overflow
mass and will power some of the mass flow away from the boundary of Roche lobe to
infinity. According to energy conservation, we have
\begin{equation}
\alpha|\dot{m}_{2}|(\phi_{\rm L1}-\phi_{\rm r1})=-(1-\alpha)|\dot{m}_{2}|\phi_{\rm L1}+L_{\rm Edd},
\label{eq_energy}
\end{equation}
where $\phi_{\rm L1}$ is the potential at the first Lagrangian point,
$\phi_{\rm r1}$ the potential at the surface of the accretor \citep{Han99}.
So the fraction of matter accreted by the neutron star is
\begin{equation}
\alpha=\frac{|\dot{m}_{2}|\phi_{\rm L1}-L_{\rm Edd}}{\dot{m}_{2}\phi_{\rm r1}}.
\label{eq_massaccretion}
\end{equation}
The potential $\phi_{\rm L1}$ and $\phi_{\rm r1}$ can be written as \citep{Kopal59}
\begin{eqnarray}
%
%
 \phi_{\rm L1} & = & -\frac{Gm_{2}}{x}-\frac{Gm_{1}}{a-x}-\frac{GM}{2a^{3}}\left(x-\frac{a}{1+q}\right)^{2},\\
\phi_{\rm r1} & = & -\frac{Gm_{2}}{a}-\frac{Gm_{1}}{\rm r1}-\frac{GM}{2a^{3}}\left[\frac{2}{3}r_{1}^{2}+\left(a-\frac{a}{1+q}\right)^{2}\right], \nonumber \\ \\
 \frac{x}{a} & = & \frac{0.696q^{1/3}-0.189q^{2/3}}{1+0.014q}.
%
%
\end{eqnarray}
The Eddington luminosity is
\begin{equation}
L_{\rm Edd}=\frac{4\pi r_{1}^{2}c\bar{g}}{\kappa},
\end{equation}
where $\kappa=0.2(1+X)$\,cm$^{2}$\,g$^{-1}$  is the opacity of the accreted
gas, $X$ the hydrogen mass fraction, and $\bar{g}$ the mean surface gravity
of the accretor. From the potential $\phi_{\rm r1}$, we obtain
\begin{equation}
\bar{g}=\frac{Gm_{1}}{r_{1}^{2}}-\frac{2}{3}\frac{GMr_{1}}{a^{3}}.
\end{equation}
Note that here we use a simplified energy conservation model to investigate
the mass change of the binary system. The actual situation may be more complicated
than this because of the involvement of other energies, such as nuclear energy and
thermal energy.
In effect, nuclear burning, thermal asymmetries and magnetic fields in accreting NSs may
contribute to GWs emission from a NS-WD binary at some level \citep{Singh20,Haskell15,Ushomirsky00,Bildsten98,Wagoner84}
which we will discuss in section \ref{sec_discussion}.
The detailed study of energy conversions is out of the scope
of this paper and deserves further investigation in future work.

\section{Results}
\label{sec_results}

\subsection{Contact frequencies}

\begin{figure}
\hspace*{-0.9cm}
\centering
\includegraphics[width=10cm,clip,angle=0]{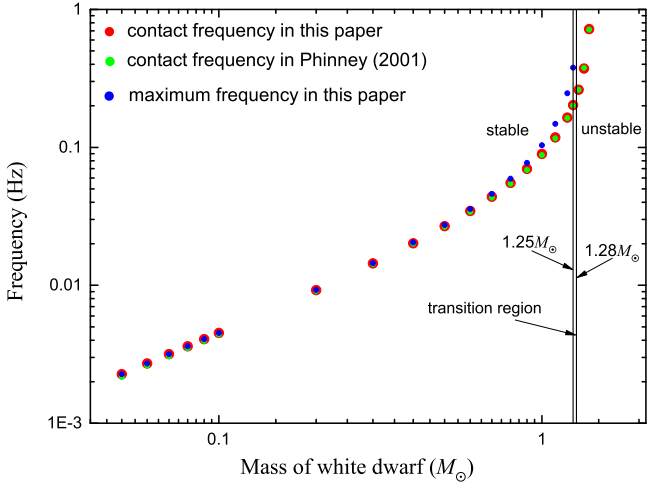}
\caption{The contact frequencies (red points) when the WD just fill its Roche lobe
and maximum frequencies (blue points) corresponding to the minimum orbital periods of
NS-WD binaries assuming $m_{\rm NS}=1.4 {\rm M}_{\odot}$) as a function of white dwarf
mass.
For comparison, green points show the contact frequencies obtained by \citet{Phinney01}.
WDs experience stable mass transfer when $m_{\rm WD}\lesssim$ 1.25 ${\rm M}_{\odot}$, and
unstable mass transfer when  $m_{\rm WD}\gtrsim  1.28  {\rm M}_{\odot}$.
There is a narrow transition region in the WD mass range 1.25 -- 1.28 ${\rm M}_{\odot}$
between stable and unstable mass transfer. }
\label{fig2}
\end{figure}

The orbital separation of a detached NS-WD binary shrinks solely due to GW radiation.
This results in the decrease of the effective Roche lobe of  the WD.
When the Roche lobe boundary reaches the WD surface, the WD starts to lose mass and the binary then becomes in ``contact''.
After contact, the radius of WD increases gradually as it loses mass.
A fraction of the lost mass may be accreted by the primary NS; additional mass may leave the binary system.
During this stage, the binary orbit evolves due to both GW radiation and mass  transfer between the two components.
We define a ``contact frequency'' of the GW radiated from the binary, which is two times the orbital frequency when the binary is just in ``contact''.

Subsequent orbital evolution of the binary depends on the competition between the angular momentum changes caused by mass transfer and GWs.
As shown in Eq.~\eqref{eq_adot}, when these processes balance, the orbital separation of a NS-WD binary reaches a minimum.
We define the corresponding GW frequency at this point as ``maximum frequency''.

Figure~\ref{fig2} shows the contact frequencies (red points) and maximum frequencies (blue points) assuming the mass of the NS component is $1.4 {\rm M}_{\odot}$.
The contact frequencies obtained here are consistent with the analytical results (green points) obtained by \citet{Phinney01}.
The numerical errors on the contact frequency obtained from our calculation are $\lesssim 0.1\%$, and the differences between our results and \citet{Phinney01} are $\lesssim 4.0\%$.

\begin{table*}
\caption{Contact frequencies (CFs) and maximum frequencies (MFs),
 WD mass-loss rates  ($\dot{m}_{\rm WD}$) at maximum frequency, and maximum WD mass-loss
rates (${\rm max}(\dot{m}_{\rm WD})$) for NS-WD binaries with $m_{\rm NS}=1.4{\rm M}_{\odot}$. The time interval from contact frequency to the given mass loss rate is shown in parentheses.
}
\hspace{-0.5mm}
\label{tabfrequency}
\begin{center}
\begin{tabular}{lcccccc} \hline
 \multicolumn{1}{c}{$m_{\rm WD}$} & \multicolumn{1}{c}{CF}& \multicolumn{1}{c}{MF} & \multicolumn{1}{c}{$\dot{m}_{\rm WD}$(time from contact)}& \multicolumn{1}{c}{${\rm max}( \dot{m}_{\rm WD})$(time from contact)}  \\
      (${\rm M}_{\odot}$)                 & (Hz)                  & (Hz)                   & (${\rm M}_{\odot}~\rm yr^{-1}$)(yrs)                & (${\rm M}_{\odot}~\rm yr^{-1}$)(yrs)                           \\
   \hline
0.05         & 0.002264     &   0.002264     & 4.82E-9 (700.0)       & 5.52E-8 (20400.0)     \\
0.06         & 0.002712     &  0.002712      & 1.05E-8 (2500.0)      & 6.30E-8 (17400.0)        \\
0.07         &  0.003165    &   0.003166       & 2.06E-8 (1850.0)   & 6.98E-8 (9060.0)     \\
0.08         & 0.003609     &  0.003610         & 4.25E-8 (700.0)       & 7.72E-8 (2700.0)      \\
0.09         & 0.004062     & 0.004063         & 6.18E-8 (542.0)        & 9.34E-8 (2365.0)        \\
0.1          &  0.004514    & 0.004515         & 7.92E-8 (470.0)      & 1.17E-7 (2030.0)          \\
0.2          &  0.009232    &  0.009248       &  1.19E-6 (218.0)      & 1.84E-6 (822.0)         \\
0.3          &  0.01440     & 0.01448         & 8.56E-6 (132.0)       & 1.36E-5 (383.0)           \\
0.4          & 0.02019      & 0.02043          & 3.78E-5 (78.6)       & 6.20E-5 (202.5)            \\
0.5         & 0.02684        & 0.02738        & 1.28E-4 (48.3)        & 2.18E-4 (110.4)             \\
0.6         & 0.03461       & 0.03571         & 3.71E-4 (30.7)        & 6.62E-4 (65.8)           \\
0.7         & 0.04391       & 0.04600         & 9.83E-4 (20.0)       & 0.00185 (39.8)             \\
0.8         & 0.05535       &  0.05923        & 0.00249 (13.1)     & 0.00498 (24.9)           \\
0.9        & 0.06998          & 0.07718           & 0.00631 (8.6)       & 0.0135 (15.2)             \\
1.0        & 0.08960           &  0.1036          &  0.0168 (5.4)    & 0.0383 (8.9)     \\
1.1        & 0.1180            &  0.1480         & 0.0513 (3.3)       & 0.119 (4.9)      \\
1.2        & 0.1645            &  0.2466        & 0.226 (1.7)      &  0.446 (2.2)      \\
1.25        & 0.2025           &  0.3791       &  0.709 (1.1)     & 1.02 (1.3)        \\
1.26        & 0.2123           &  0.4323      &    0.976 (1.0)       &   1.25 (1.2)       \\
1.27        & 0.2225           &   0.5226      &  1.47 (1.0)         &    1.60 (1.0)               \\
1.28        & 0.2341           &   unstable         &  ---         &      ---               \\
1.3        & 0.2616            &   unstable         &   ---       &    ---       \\
1.4        & 0.7194            &   unstable         &    ---       &   ---     \\
\hline
\hline
\end{tabular}
\end{center}
\end{table*}

\subsection{Stability of Roche lobe overflow}

The condition for stable Roche lobe overflow is considered to be when
the first order change of radius in logarithm with respect to the mass change of
secondary (mass donor) is smaller than the first order change
of the effective Roche lobe radius with respect to the mass change,
i.e. $\frac{\partial \ln r_{2}}{\partial \ln m_{2}} < \frac{\partial \ln r_{\rm 2L}}{\partial \ln m_{2}}$.
If $\frac{\partial \ln r_{2}}{\partial \ln m_{2}} > \frac{\partial \ln r_{\rm 2L}}{\partial \ln m_{2}}$,
the increase in WD radius after loss of mass is larger than the effective expansion of the Roche lobe.
This will further aggravate the mass loss and the Roche lobe overflow tends to be unstable.
Otherwise, after the loss of mass, the secondary (mass donating star) tends to be confined
in its Roche lobe, the mass loss is reduced and the Roche lobe overflow tends to be stable.

We do not use this condition to distinguish the stable and unstable Roche lobe overflow in our numerical calculations.
Instead, the WD mass loss rate $\dot{m}_{2}$ is calculated
according to Eq.~\eqref{eq_mtrate}, which explicitly illustrates the dependency between the
mass loss rate and the change of WD radius and its Roche lobe radius with respect to time.
We can  understand if the Roche lobe overflow is stable or not in terms of the second
order derivative of WD mass with respect to time, i.e. $\ddot{m}_{2}$.
If $\ddot{m}_{2}>0$, $\dot{m}_{2}$ increases, the WD will lose more and more mass, and the Roche lobe overflow is unstable.

Our results indicate that WDs experience stable Roche lobe overflow when their masses are
$\lesssim$ 1.25 ${\rm M}_{\odot}$.
The consequence of  stable Roche lobe overflow in our calculations
is that the WD  mass loss rate increases from 0 to a peak value and then declines to
a limiting rate which will be derived in \S\,3.5.
During the mass loss process, the WD  will first spiral in until the orbital separation  reaches a minimum,  and the GW frequency reaches a maximum.
Then, as mass transfer dominates over GWs, the orbital separation
starts to increase and the corresponding GW frequency decreases.

When $m_{\rm WD}\gtrsim 1.28 {\rm M}_{\odot}$, WDs experience unstable Roche lobe
overflow, which results in fast spiral-in until the orbital separation
is reduced to equal the sum of WD and NS radius. The WD will be disrupted just around this point, producing a GW burst.
In the narrow range $m_{\rm WD}\ \sim 1.25-1.28 {\rm M}_{\odot}$,
the Roche lobe overflow can be stable, but the WD radius is very close to that of the
orbital separation.
This implies that NS may sweep through the edge of the WD quickly in this narrow mass range.
The mass boundary weakly depends on the NS mass. When changing the NS mass to 3.0 $\rm M_{\odot}$, the transition region of WD mass
becomes $m_{\rm WD}\ \sim 1.23-1.24 {\rm M}_{\odot}$.
The shape of white dwarf is likely to be influenced by tidal effects;  the consequences of this tidal deformation will be discussed in \S\,\ref{sec_tidal_deformation}.

In Table~\ref{tabfrequency}, we list the contact frequencies (CFs), maximum frequencies (MFs), mass loss rates of WDs at the
maximum frequencies ($\dot{m}_{\rm WD}$), maximum mass loss rates (${\rm max}( \dot{m}_{\rm WD})$),
and the time interval from contact frequencies to the mass loss rates with unit of yrs, respectively, as a function of WD mass.
This, in a marginally unstable case, a WD with initial mass of $1.27 {\rm M}_{\odot}$ will lose $\sim 0.103  {\rm M}_{\odot}$ in $24.5$\,days.
However, a WD with initial mass $0.06 {\rm M}_{\odot}$ takes about $14\,900$\,yrs from
maximum frequency to maximum mass loss rate, and only loses a small amount of mass, about $5.5\times 10^{-4} {\rm M}_{\odot}$.

\subsection{Conservative and non-conservative evolution}

The adopted criterion for mass transfer in NS-WD binaries to be conservative or non-conservative is the Eddington accretion rate ($\approx6.0\times10^{-8} {\rm M}_{\odot}{\rm yr}^{-1}$), derived from the Eddington accretion luminosity.
If the accretion luminosity is less than the Eddington limit $L_{\rm Edd}$, mass transfer is conservative, which means all matter lost from the WD is accreted by the NS.
If the accretion luminosity is larger than $L_{\rm Edd}$ and the diffusion timescale of outward photons released in the accretion flow is slower than the timescale of photons swept inward by the accretion flow itself, the excess fraction of energy is absorbed in the accretion flow and will accelerate some of the accreted matter away
from the boundary of the Roche lobe to infinity.
Here we do not take into account the influence of the disk wind, which may have some effects on the orbital evolution.
Note that the mass loss caused by the wind is not observationally determined for
4U\,1820--30.
For other low-mass binaries, such as GX\,13+1, it is estimated to be $\sim(0.6-6)\times10^{-8}$ ${\rm M}_{\odot}~\rm yr^{-1}$ \citep{Allen18}.
The mass lost via the disk wind is less than that by Roche lobe overflow
($\sim6 \times 10^{-8}$ ${\rm M}_{\odot}~\rm yr^{-1}$, see next section)
assuming that 4U\,1820--30 has a similar wind to GX\,13+1.
This will result in a less significant orbital change for 4U\,1820--30 than by Roche lobe overflow alone.
We neglect the influence of wind on the orbital change in this paper and defer a detailed investigation to a future investigation.

Our calculations indicate that, for a primary NS of $1.4 {\rm M}_{\odot}$, when $m_{\rm WD}<0.058 {\rm M}_{\odot}$,
the mass transfer is fully conservative, and when $m_{\rm WD}\gtrsim 0.058 {\rm M}_{\odot}$, the mass transfer becomes non-conservative.
Considering $m_{\rm WD}\simeq 0.1 {\rm M}_{\odot}$, when the mass loss rate reaches a maximum value, $\approx 51\%$ of the lost mass
is accreted by the NS, otherwise, $> 51\%$ of the lost mass  is accreted.
For $m_{\rm WD} = 1.0 {\rm M}_{\odot}$ and $1.2 {\rm M}_{\odot}$, only $\sim0.1\%$ and $\sim0.074\%$ of the mass loss can be
accreted when the WD mass loss rate reaches maximum.

In the unstable mass transfer case, NS-WD binaries usually have high chirp mass and close orbit separation,
which results in the GW induced orbital separation change always being larger than the change caused by the mass transfer.
These binaries eventually become runaway mergers.
The WD mass loss rates increase nonlinearly, leading to a large fraction of mass being lost from the system.
For WDs with masses of $1.28 {\rm M}_{\odot}$ and $1.4 {\rm M}_{\odot}$,
$\lesssim0.47\%$ and $\lesssim1.0\%$ of the lost mass can be accreted by the NS.

\subsection{Evolution of NS-WD binaries}

\begin{figure}
\hspace*{-0.7cm}
\centering
\includegraphics[width=9cm,clip,angle=0]{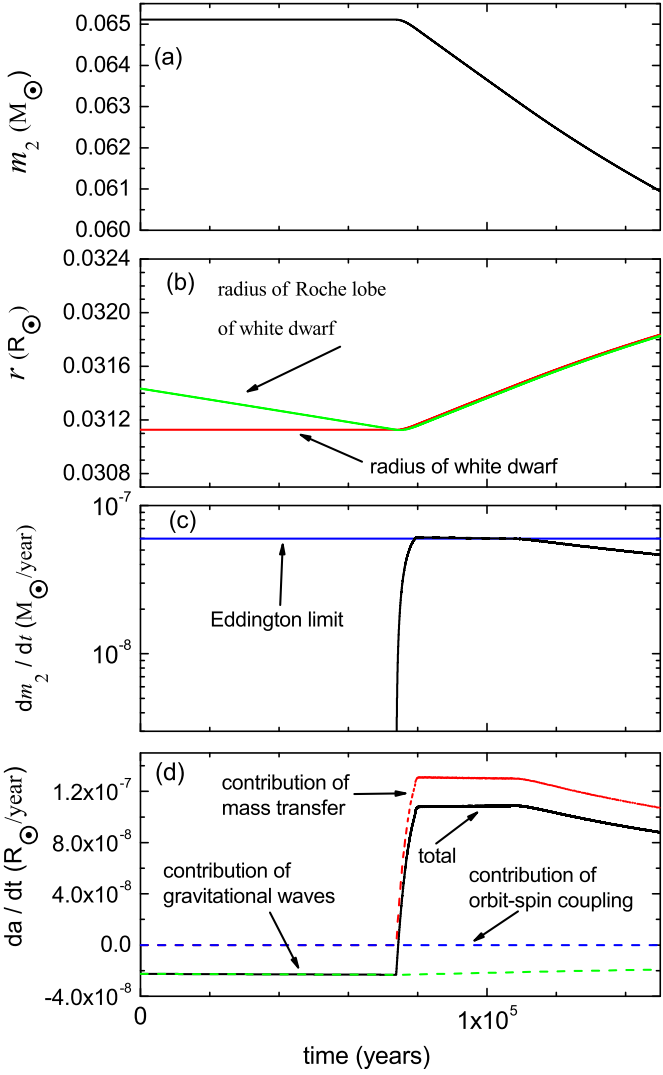}
\caption{The time-dependent variation of physical parameters of the WD companion of ultracompact X-ray binary 4U\,1820--30 ($1.0  {\rm M}_{\odot}$ NS + $0.06511 {\rm M}_{\odot}$ WD binary). Panels (a): WD mass; (b) WD radius and Roche lobe radius; (c)  WD mass loss rate; (d) change rate of semimajor axis (including contributions from GWs, mass transfer, and orbit-spin coupling).}
 \label{fig3}
\end{figure}

\begin{figure}
\hspace*{-0.4cm}
\centering
\includegraphics[width=9cm,clip,angle=0]{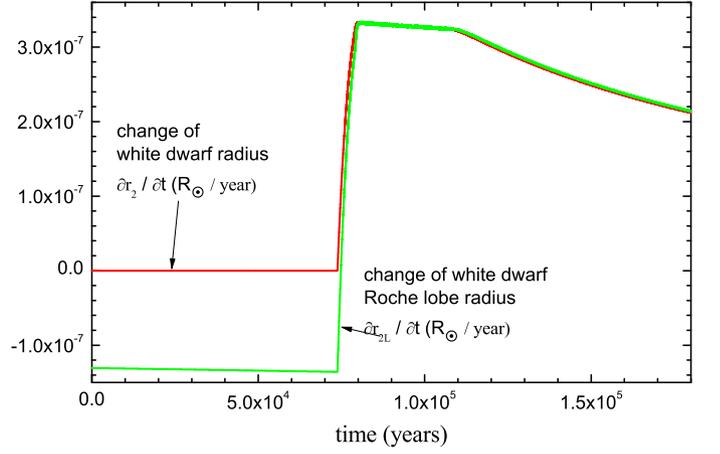}
\caption{The first order derivatives of white dwarf radius and Roche lobe radius with respect to time, for the same system as in Fig.~\ref{fig3}.}
 \label{fig4}
\end{figure}

Figure~\ref{fig3} shows the time-dependent changes of physical parameters of a NS-WD binary with masses $1 M_\odot+0.0651 {\rm M}_{\odot}$, which could represent the evolution of the ultracompact X-ray source 4U\,1820--30.
Panels include (a): WD mass; (b): WD radius and Roche lobe radius (c): WD mass loss rate (including Eddington accretion rate); and (d) change rate of the binary semimajor axis.
Panel (d) implies that the orbital separation gradually
decreases as a consequence of GW radiation removing energy and angular momentum from the binary system.
When the effective Roche lobe radius ($r_{2L}$) reaches the WD radius ($r_{\rm 2}$),  such that $r_{2L}=r_{\rm 2}$, significant mass transfer commences.
A fraction of mass from the WD will be accreted by the NS through the region near L1.
The mass loss makes the radius of the WD increase continuously and the mass ratio $m_2/m_1$ decreases.

When the first derivative of the WD radius is equal to that of the WD effective Roche lobe radius with respect to time and mass, the mass transfer tends to be stable, as shown in Figs.~\ref{fig3} and \ref{fig4}.
The maximum mass loss rate of WD is about $6.10 \times10^{-8} {\rm M}_{\odot} {\rm yr}^{-1}$,
which is slightly greater than the Eddington accretion rate $\sim 6.0 \times10^{-8} {\rm M}_{\odot} {\rm yr}^{-1}$.
This means that the mass transfer between the two stars is non-conservative
after a time interval of $28\,176$\,yrs (the minimum value of $\alpha$ is $\sim 0.98$), and most of the mass lost from the WD is accreted onto the NS.
If we assume that the release of gravitational potential energy caused by accretion is completely converted to X-ray luminosity (radiation efficiency $100\%$), then  $L_{\rm X}=\frac{Gm_{1}\alpha \dot{m}_{2}}{r_{1}}$.
The X-ray luminosity produced by the accretion rate corresponding to the highest mass loss rate of the white dwarf is about $2.5 \times10^{38} {\rm ergs\,s}^{-1}$, or about 4 times larger than the observed value $6 \times10^{37}{\rm ergs\,s}^{-1}$ \citep{Stella87}.

Figure~\ref{fig4} shows the change of WD radius $\partial r_{2}/\partial t$ and the effective Roche lobe radius
$\partial r_{\rm 2L}/\partial t$.
It can be seen that the mass transfer is unstable at the onset of Roche lobe overflow, and that the mass loss rate increases dramatically (Fig.~\ref{fig3}).
When $\partial r_{2}/\partial t=\partial r_{\rm 2L}/\partial t$, the mass loss rate approximately reaches the maximum.
After that, the change of effective Roche lobe $\partial r_{\rm 2L}/\partial t$ is slightly greater than the change of WD radius $\partial r_{2}/\partial t$, resulting in slow and approximately linear decrease of mass loss rate.

\begin{figure}
\hspace*{-0.5cm}
\centering
\includegraphics[width=9cm,clip,angle=0]{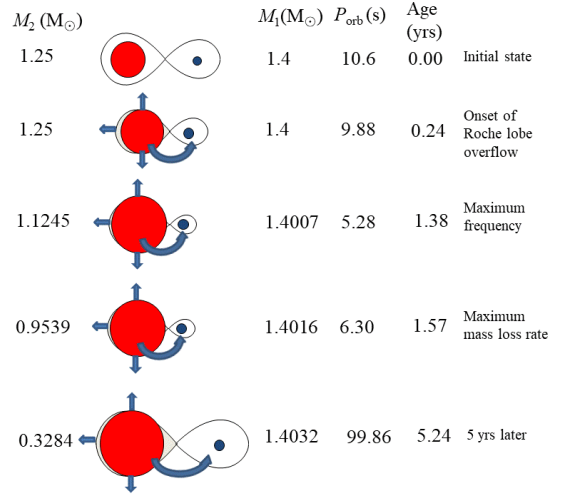}
\caption{Schematic showing the orbital evolution of an NS-WD binary with initial masses of $1.4M_\odot+1.25 {\rm M}_{\odot}$. Rows from top to bottom represent different binary stages, i.e., the initial binary, first contact, period minimum (or  maximum GW frequency), maximum mass loss rate, and $5$\,yrs after contact, respectively.
}
 \label{fig5}
\end{figure}

\begin{figure}
\hspace*{-0.5cm}
\centering
\includegraphics[width=10cm,clip,angle=0]{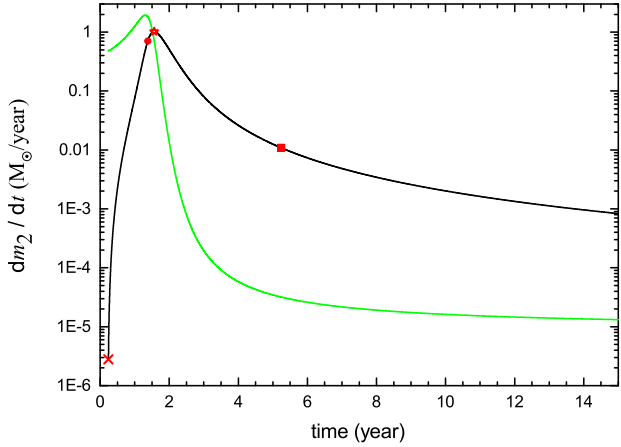}
\caption{The mass-loss rate of the WD in a NS-WD binary with initial masses  $(1.4+1.25){\rm M}_{\odot}$ is shown as black line.
The green line show the mass loss rate which balances the change of orbital separation contributed from gravitational waves, calculated by Eq.\,\ref{eq_balancemdot}.
The red cross, solid circle, star and solid square
show the points when the white dwarf fills its Roche lobe, the gravitational wave frequency reaches its
maximum value, the mass loss rate reaches the peak value, and 5 years later after first contact,
respectively.}
 \label{fig6}
\end{figure}

Figure~\ref{fig5} depicts the orbital evolution of a high mass NS-WD binary with initial masses of $1.4M_\odot+1.25{\rm M}_{\odot}$ and orbital period of $10.6$\,s.
The WD fills its Roche lobe after $0.24$\,yrs when the orbital period is $\sim 9.88$\,s (corresponding to GW frequency $\sim 0.20$\,Hz).
After $1.14$\,yrs, the binary reaches the maximum GW frequency $\sim0.38$\,Hz, and the WD loses $\sim10\%$ of its initial mass. After $0.186$\,yrs, the mass loss rate of WD reaches its maximum which is about $1.02 {\rm M}_{\odot} {\rm yr}^{-1}$, and the GW frequency
now becomes $\sim0.32$\,Hz.
About $5$\,yrs after first contact, the white dwarf has lost $73.7\%$ of its initial mass and the binary orbital period increases to $\sim99.86$ s.
The mass loss rate of WD in this binary is shown in Fig.~\ref{fig6}.
In Section~\ref{lab_gwradiation}, we will show that the mass transfer process for contact NS-WD binaries may be probed by space-based GW detectors and constrained by GW signals.

\subsection{Evolution of orbital period and GW frequency}

\begin{figure}
\hspace*{-0.4cm}
\centering
\includegraphics[width=9cm,clip,angle=0]{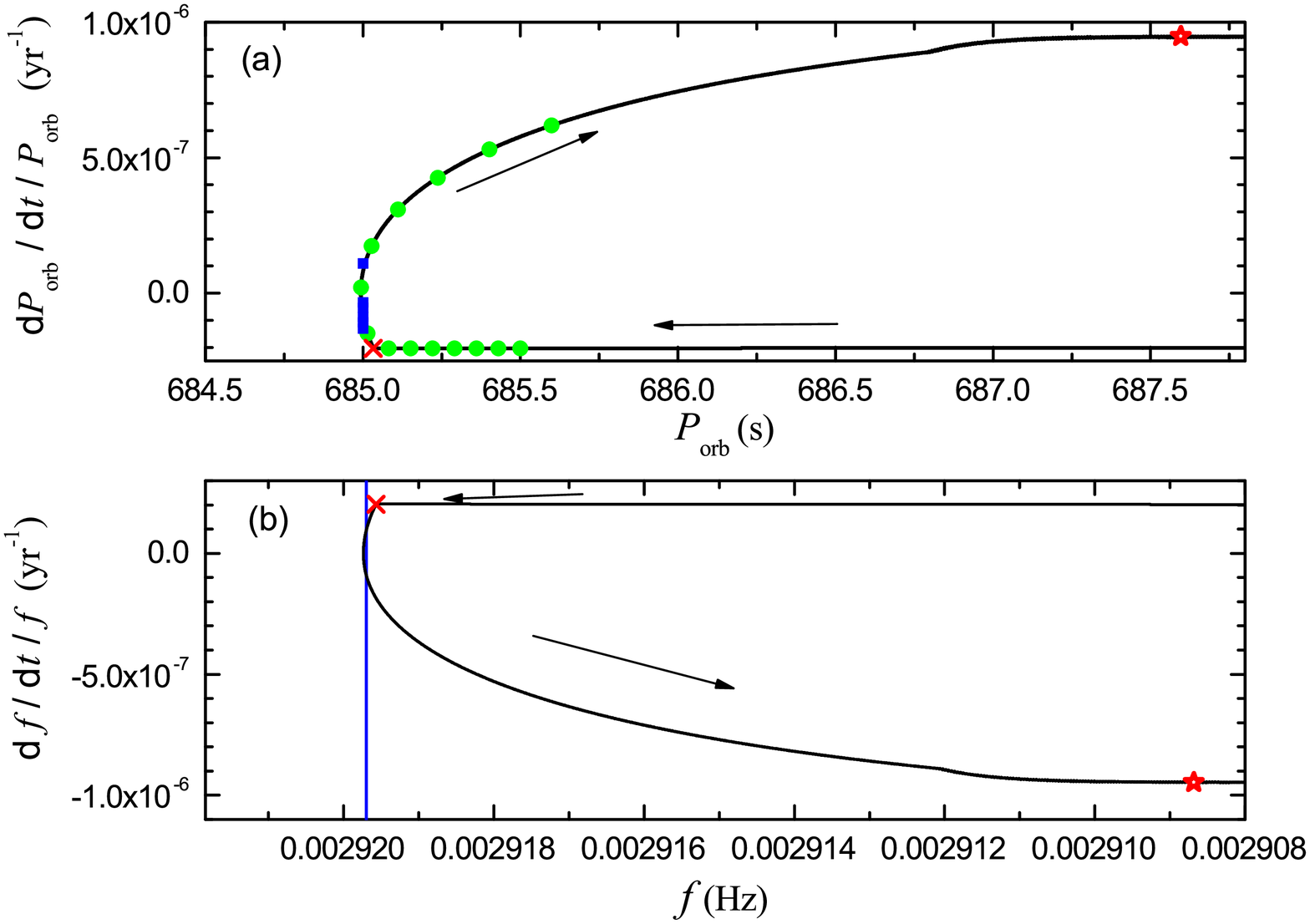}
\caption{Panel (a): the relative rate of change of orbital period ($\dot{P}/P$) of a $1.0 {\rm M}_{\odot}$) NS + $0.06511 {\rm M}_{\odot}$ WD binary as a function of  orbital period.
The period of 4U\,1820--30 is shown as a black line.
Blue squares denote the observations of UCXB 4U\,1820--30 (\citealt[][]{Peuten14}; \citealt{vanderKlis93}a; \citealt{vanderKlis93b}b; \citealt{Tan91, Morgan88, Chou01}). Black arrows show the direction of evolution.
Green points are selected from our calculations, and the time interval between neighbouring points is $10^{3}$\,yrs.
Panel (b): the relative rate of change of the binary GW frequency as a function of the GW frequency (black line).  Since $f_{\rm GW} = 2 / P_{\rm orb}$, $\dot{f}/f \equiv - \dot{P}/P$.
The blue line shows the current frequency ($2/685$\,Hz) of 4U\,1820--30.
Black arrows show the direction of evolution.
Red crosses and stars in both (a) and (b) show the points when the  WD fills its Roche lobe and its mass loss rate reaches the peak value,
respectively.  The NS-WD binary is assumed to be in a circular orbit.
}
 \label{fig7}
\end{figure}

\begin{figure}
\hspace*{-0.4cm}
\includegraphics[width=9cm,clip,angle=0]{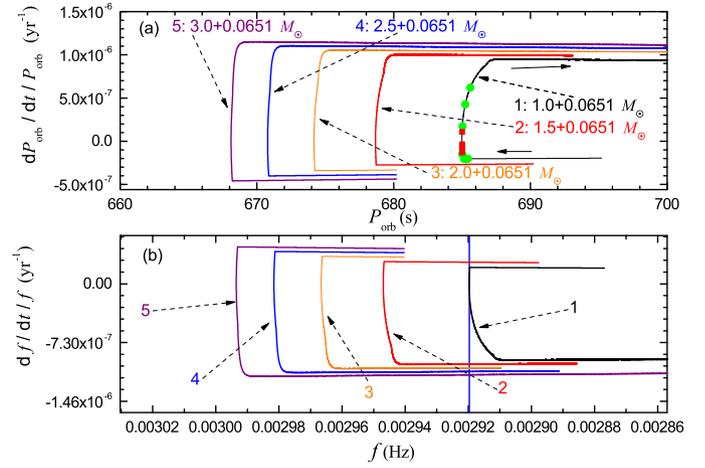}
\caption{Panel(a): As Fig.~\ref{fig7}a for additional NS-WD models (denoted by numbers 1 to 5, colored solid line).
The case for 4U\,1820--30 is shown as a black line.
Panel (b): As Fig.~\ref{fig7}b for additional NS-WD models.  The vertical blue line denotes the current frequency (2/685 Hz) of 4U\,1820--30.}
 \label{fig8}
\end{figure}

\begin{figure}
\hspace*{-0.7cm}
\centering
\includegraphics[width=9cm,clip,angle=0]{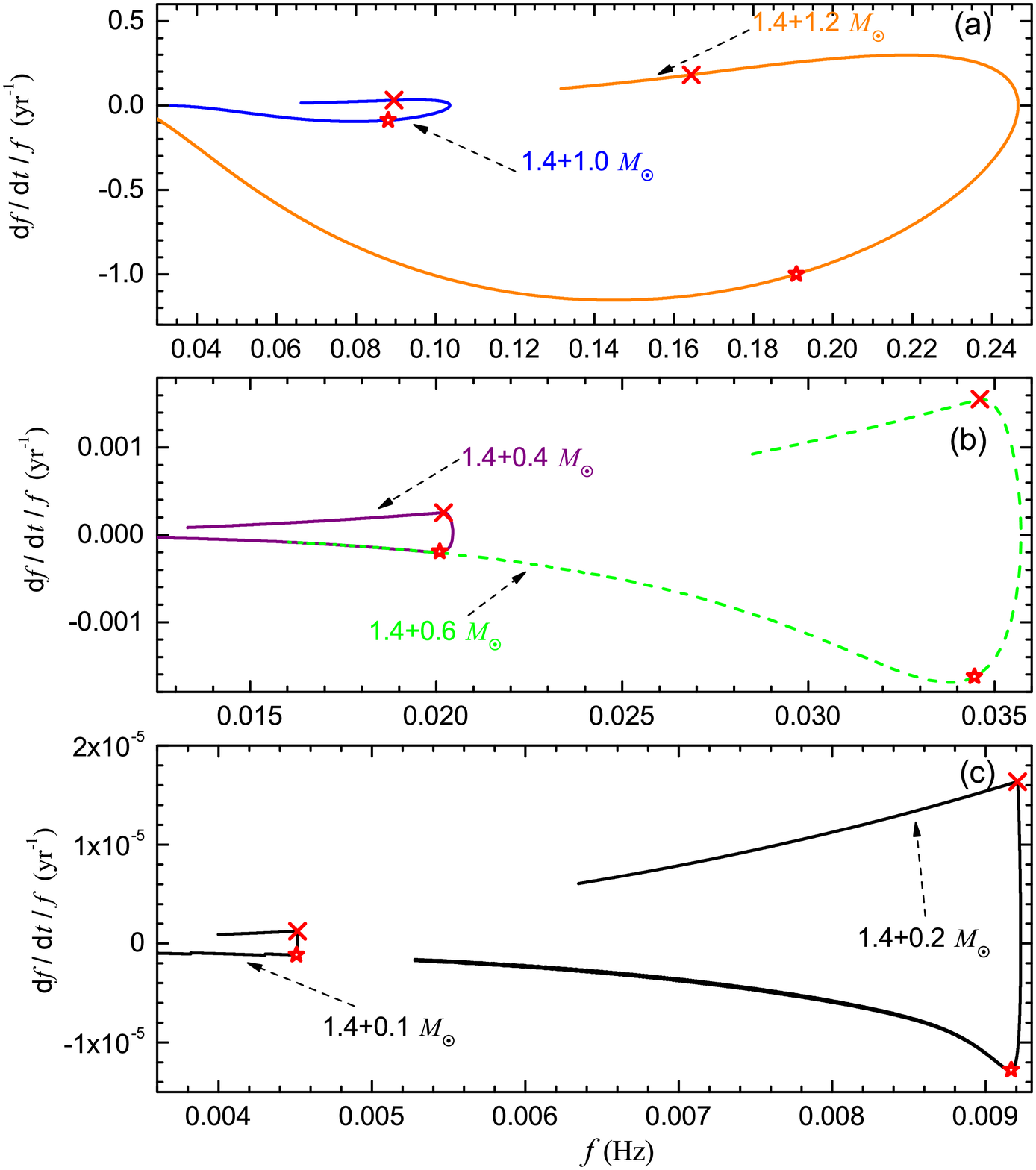}
\caption{
 Relative change rate of GW frequencies of some NS-WD binaries as a function of the GW frequency. Panels (a), (b), and (c) show the evolution paths for NS-WD binaries with masses ($1.4M_\odot + 1.0M_\odot$ and $1.4M_\odot+1.2M_\odot$), ($1.4M_\odot + 0.6M_\odot$ and $1.4M_\odot+0.4M_\odot$), and ($1.4M_\odot + 0.2M_\odot$ and $1.4M_\odot+0.1M_\odot$), respectively.
Red crosses and stars show
the points when the WD companion fills its Roche lobe and its mass loss rate reaches the peak value,
respectively.
}
 \label{fig9}
\end{figure}

Figure~\ref{fig7} shows the relative change rate of the orbital periods (panel (a))
and GW frequency (panel (b)) of 4U\,1820--30 as a function of its period,
assuming that 4U\,1820--30 is a $1.0 {\rm M}_{\odot} + 0.06511 {\rm M}_{\odot}$ NS-WD binary.
When the angular momentum change caused by GWs is balanced by other
factors, the change of orbital separation will be zero.
For simplicity, if we assume that the mass loss effect is the dominant factor to balance the
GW decay, we obtain the relation between mass transfer rate and orbital parameters as
\begin{equation}
\begin{split}
&\dot{m}_{2}=-C\frac{32G^{3}m_{1}m_{2}^{2}M}{5c^{5}a^{4}}\\
&~~~~=C\frac{4\pi^{2}}{\sqrt{5}c}m_{2}f^{2}hR_{\rm b},\\
&C=\left[1 - \alpha q
- (\beta+\frac{1}{2})(1-\alpha)\frac{q}{1+q}
- \alpha \sqrt{(1+q)r_{\rm h}}  \right ]^{-1},
\end{split}
\label{eq_balancemdot}
\end{equation}
where $M=m_{1}+m_{2}$ is the total mass of a binary, $h$ is the averaged GW strain amplitude over one orbital period, $R_{\rm b}$ is the distance from the Earth, and $C$ is a dimensionless coefficient having value $\sim 2.91-1.35$ when the WD mass is in the range of $0.05-1.25 {\rm M}_{\odot}$.

Note that the mass transfer rate still increases when the change of orbital separation is greater than zero until the mass loss rate reaches the largest value, and at this point, the change of the WD radius is approximately equal to the change of its effective Roche radius.

Our results show that the evolution of  NS-WD binaries can be divided into the following four stages.
(1) The orbital period decreases because GW radiation dominates the energy and angular momentum evolution of the binary.
(2) The orbital period continues to decrease, but $\dot{P} \rightarrow 0$ as significant mass transfer participates in the angular momentum evolution to cancel the GW effect. When $\dot{P} = 0$, the orbital period reaches its minimum value.
(3) The orbital period increases because mass transfer dominates the energy and angular momentum evolution.
(4) The orbital period continues to increase but its first-order change goes down slowly.
i.e., $\dot{P}>0$, $\ddot{P}<0$ and $\ddot{P}\rightarrow 0$.
The NS-WD binary continues to evolve until the WD cannot provide sufficient mass to sustain mass transfer.
We here assume that there are no other physical processes that result in a sudden change of the mass transfer rate.
We show a number of evolution tracks for NS-WD binaries with different masses in Fig.~\ref{fig8}.

Most observations of 4U\,1820--30 indicate $\dot{P} < 0$ (e.g. \citealt[][]{Peuten14}; \citealt{vanderKlis93}a; \citealt{vanderKlis93b}b; \citealt{Tan91, Morgan88, Chou01}).
implying that this NS-WD binary is experiencing stage (2) evolution.
In this case, the corresponding GW frequency change rate would be $\dot{f}/f \sim9.3 \times 10^{-8}  {\rm yr}^{-1}$
(GW frequency $0.0029197\,{\rm Hz} = 2 / (685\,{\rm s})$).
However, some models predicted a positive period change, e.g., $\dot{P}_{\rm orb}/P_{\rm orb}\approx (1.1\pm0.5) \times 10^{-7}$ yr$^{-1}$ \citep{Rappaport87}, which suggests that this binary might be at the third stage.
In this case, the corresponding GW frequency change rate would be $\dot{f}/f \sim -1.1 \times10^{-7} {\rm yr}^{-1}$.

To facilitate comparison, Figs.~\ref{fig8} and \ref{fig9} show the evolution of NS-WD binaries with different masses, assuming the upper mass limit of NS is about $3.0 {\rm M}_{\odot}$ (Fig.~\ref{fig8}) and that of WD is about $1.4 {\rm M}_{\odot}$ (Fig.~\ref{fig9}), respectively.
It can be seen that the larger the mass of the NS component, the greater the absolute value of $\dot{P}$  at the onset of mass transfer.
Fig.~\ref{fig9} implies that the orbital parameters, mass transfer rate, and evolution stage of NS-WD binaries may be identified by measuring the GW frequency and its rate of change, i.e. the detection of gravitational waves in the future can help us to find the high-mass compact binaries.

By combining Kepler's law, Eq.\,\ref{eq_adot} and the orbital period-GW frequency relation $f=2/P_{\rm orb}$,
and by assuming that the NS-WD orbit binary is already circularized by tidal interaction and gravitational radiation,
an analytical solution for the conservative evolution ($\alpha=1$) of GW frequency ($\dot{f}/f$) of a contact NS-WD binary
can be obtained:
\begin{equation}
\begin{split}
\frac{\dot{f}}{f}=&\frac{96}{5}\pi^{8/3}c^{-5}G^{5/3}m_{1}m_{2}M^{-1/3}f^{8/3}+ 3C^{-1}\frac{\dot{m}_{2}}{m_{2}}\\
&-\frac{3km_{1}r_{1}^{2}}{\tau_{\rm s}J_{\rm orb}}(\omega_{\rm s}-\omega_{\rm o}).
\label{eq_gwf}
\end{split}
\end{equation}
The three terms on the right hand side  represent the contribution of GWs,
mass transfer and orbit-spin coupling to the GW frequency evolution $\dot{f}/f$.
If $\dot{f}/f$, $m_{1}$ and $m_{2}$ can be determined via measurements of GW and electromagnetic waves
from a NS-WD binary, we should be able to deduce the mass transfer $\dot{m}_{2}$.
Equation\,\ref{eq_gwf} explicitly describes $\dot{f}$ in terms of  mass transfer and orbital parameters, in contrast to the dynamical chirpmass used to describe $\dot{f}$ by \citet{Tauris18}
(cf. Eq. (4) in their paper).

\subsection{Effect of the orbit-spin coupling}
\label{lab_oscoupling}

The coupling between the orbital angular momentum of the NS-WD binary and the spin of the primary NS may influence the evolution of the system angular momentum, depending on the synchronization time scale, initial orbital period, and spin parameters.
From the third term in the r.h.s. of Eq.~\ref{eq_adot}, i.e., the change of semimajor axis caused by the orbit-spin coupling, it can be seen that when the spin angular velocity is greater than the orbital angular velocity, the orbital separation increases.
Otherwise, the semimajor axis decreases. Because the angular frequency of pulsars is usually higher than $0.1$\,Hz, for UCXBs the spin  angular velocity is generally higher than the orbit angular velocity, and the orbit-spin coupling can lead to an increase of the orbital semimajor axis.
However, the magnitude of the change ($\dot{a}$) is greatly affected by the synchronization timescale, which is closely related to the tidal interaction, magnetic field, and energy dissipation.

For 4U\,1820--30, if we take the masses of the  NS and  WD  to be $m_{1}=1.0 {\rm M}_{\odot}$ and $m_{2}=0.0651 {\rm M}_{\odot}$, respectively, and with radii $r_{1}=1.44\times 10^{-5} {\rm R}_{\odot}$ and $r_{2}=0.031127 {\rm R}_{\odot}$, orbital semimajor axis $a=0.1708 {\rm R}_{\odot}$, rotation frequency $f_{\rm s}=\omega_{\rm s}/2\pi=1$ Hz and magnetic field $B=10^{9}$\,G, the synchronization timescale is $\tau_{\rm s}\approx (4-5)\times10^{10}$\,yrs. 
When the mass transfer of the NS-WD binary is nearly stable, the contribution of the orbit-spin coupling to the change of semimajor axis is about $2.3\times10^{-12} {\rm R}_{\odot}~{\rm yr}^{-1}$, or about $4-5$ orders of magnitude smaller than the contributions from the mass transfer and GWs ($1.2\times10^{-7} {\rm R}_{\odot}~{\rm yr}^{-1}$ and $1.7\times10^{-8} {\rm R}_{\odot} ~{\rm yr}^{-1}$). 
In order to see the effect of orbit-spin coupling, we compute a grid of $f_{\rm s}=0.1, 10, 1000$ Hz and $\log(B/\rm G)=8, 10, 12$ for this binary
and another binary with higher masses of $(1.4+1.2) {\rm M}_{\odot}$. Our results indicate that the orbit-spin coupling becomes stronger with increasing
rotation frequency and magnetic field. However, even with $f_{\rm s}=1000$ Hz and $\log(B/\rm G)=12$, the effect of orbit-spin coupling on
the orbital separation is less than that of mass transfer by a factor of $\sim$8 when the mass transfer and gravitational waves is nearly in equilibrium.
For the $(1.4+1.2) {\rm M}_{\odot}$ NS-WD binary with $f_{\rm s}=1000$ Hz and $\log(B/\rm G)=12$, the synchronization timescale gradually decreases from
the beginning of mass transfer until the orbital separation reach its minimum, and thereafter the timescale constantly increases mainly due to the change of
orbital separation. When the effect of orbit-spin coupling is the greatest, it is less than that of mass transfer by a factor of $\sim$600, most likely because of
the large mass loss rate in the high mass WD-NS binary.
The effect of orbit-spin coupling will be discussed in section \ref{lab_gwradiation}. 

Estimates according to Eq.~\ref{eq_stid} indicate that the synchronization timescale due to tidal interaction is quite
long ($\tau_{\rm s,tid}\gtrsim10^{9}$\,yr) and the corresponding orbit semimajor axis change $\lesssim10^{-11} R_{\odot} {\rm yr}^{-1}$.
These results show that the influence of the orbit-spin coupling should be considered
when the NS has strong magnetic field, fast rotation frequency or the WD has high luminosity (see section \ref{sec_method} for some calculations).
Time-dependent evolution models of the magnetic field, rotation frequency and luminosity are needed for the effect of orbit-spin coupling on
the GW signal of NS-WD binaries in future work.

\subsection{Influence of tidal deformation}
\label{sec_tidal_deformation}

Tidal interaction can deform the WD because of strong gravitational potential.
We refer to the shape deformation as the tidal bulge.
The height of the bulge referred to the mean radius is about \citep{goldreich66}:
\begin{equation}
\begin{split}
&\Delta r_{2}=\epsilon\frac{3}{4}\frac{m_{1}r_{2}^{4}}{m_{2}a^{3}},\\
&\epsilon=\frac{5}{2+19\iota_{2}/g_{2s}\rho_{\rm 2s} r_{2}},
\label{eq_bulgeheighttide}
\end{split}
\end{equation}
where $\epsilon$ is the correction factor for the rigidity of the white dwarf and
for a second degree disturbance of the tidal potential due to the deformation.
$\iota_{2}$, $g_{2s}$ and $\rho_{\rm 2s}$ represent the rigidity, surface
gravity, and tidal bulge density of a white dwarf respectively \citep{goldreich66}.
For WDs, we assume $19\iota_{2}\lesssim g_{2s}\rho_{\rm 2s} r_{2}$ which gives the maximum value
$\epsilon_{\rm max}=5/2$.
In order to simplify our model in this paper, we take $\epsilon=\epsilon_{\rm max}$ as a constant
to reduce the number of free parameters.
In due course, the rigidity parameter $\iota_{2}$ and its dependence on the interior physics of the white dwarf needs to be investigated in more detail.

When taking  tidal interaction into account, the (1.0+0.065)${\rm M}_{\odot}$
NS-WD model for 4U\,1820--30 is no longer applicable, since the tidal interaction will change the contact frequency.
Our computation indicates the NS-WD model would need masses (1.0+0.078)${\rm M}_{\odot}$ in order to explain the observations.
This model gives a binary chirpmass of $\sim0.21$ $M_{\odot}$.
In contrast, the binary chirpmass is $\sim0.19$ $M_{\odot}$ without tidal interaction.
This suggests the effect of tidal interaction in NS-WD binary evolution might be detectable in future GW observations.

For NS-WD binaries with larger masses, e.g., $(1.4+1.25){\rm M}_{\odot}$, the white dwarf radius without (with)  tidal deformation $r_{2}\thickapprox0.00507$ $R_{\odot}$ (0.00564 $R_{\odot}$),
the orbital separation $a=0.0137$ $R_{\odot}$ (0.0139 $R_{\odot}$),
and the orbital period $P_{\rm orb}\thickapprox9.88~s$ ($10.07~s$)  at the onset of mass transfer.
Approximately 1.14 yrs (0.48 yrs) after contact without (with) consideration of tidal interaction (when the GW frequency reaches maximum), the NS-WD masses become $1.4007+1.125 {\rm M}_{\odot}$ ($1.4002+1.2202 {\rm M}_{\odot}$),
the white dwarf radius becomes $0.0067 R_{\odot}$ ($0.0065 R_{\odot}$), and the orbital separation is about $0.0089 R_{\odot}$ ($0.0128 R_{\odot}$).
The minimum orbital period $P_{\rm orb}\thickapprox5.28~s$ ($\sim8.91~s$).

Our results indicate that when initial white dwarf mass is $\lesssim1.25~{\rm M}_{\odot}$, binary orbital
separation is always greater than the sum of white dwarf radius and neutron star radius.
The influence of tidal deformation on the mass boundary of stable and unstable mass transfer is small,
so we set a binary of $(1.4+1.25)~{\rm M}_{\odot}$ as a tentative example for tests of future observations.
we only present our model with consideration of the tidal deformation in this section because the tidal deformation in NS-WD evolution remains unclear and a simplified model may help us understand the future GW observations.

\subsection{GW signals from NS-WD binaries}
\label{lab_gwradiation}

\begin{figure}
\hspace*{-1.1cm}
\centering
\includegraphics[width=10cm,clip,angle=0]{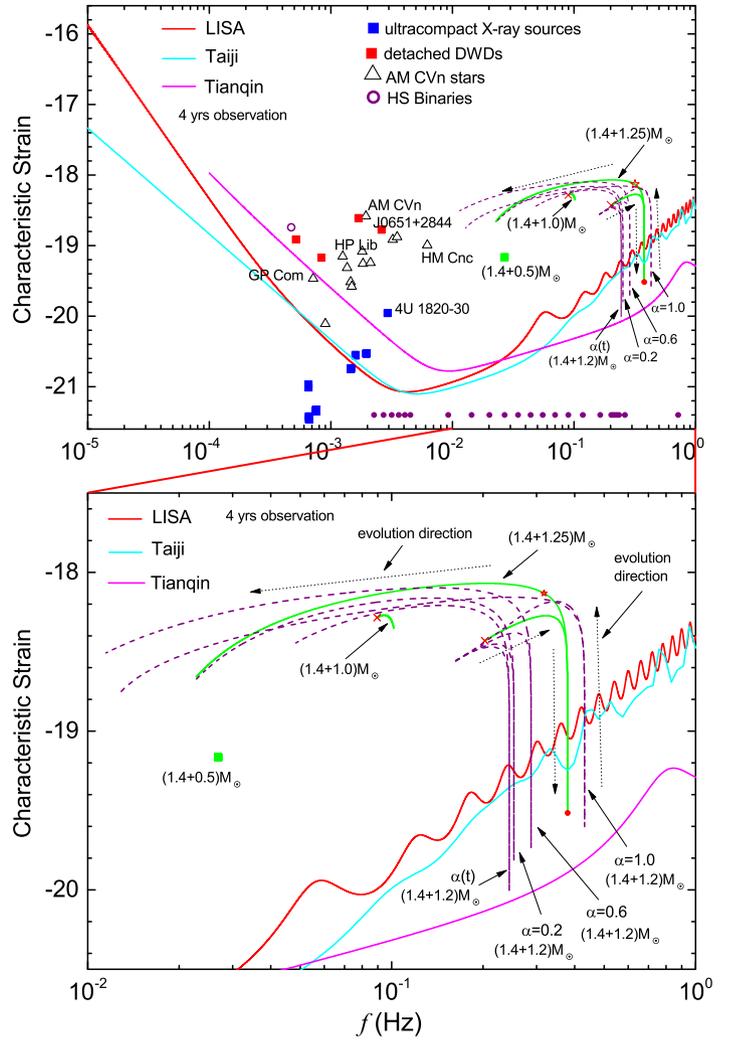}
\caption{Top panel: GW characteristic strains of example NS-WD binaries as a function of frequency (assuming a 4 year observation period).
For comparison, we also plot the characteristic strain of other GW sources, such as detached double WDs, AM CVn stars (including some double WD X-ray sources),
and hot subdwarf (HS) binaries.
Green squares and lines show the evolution track of contact NS-WD binaries with masses of (1.4 + 0.5, 1.0, 1.25) $M_\odot$, respectively.
The red crosses on the two green lines denote the onset of the mass transfer. 
Purple dashed lines show the evolution of (1.4+1.2) $M_\odot$ binary with $\alpha$ being the accretion parameter in Eq.\,\ref{eq_adot}.
$\alpha(t)$ is the time-dependent parameter in our model.
Red solid circle and open star mark the points where the GW frequency reaches its maximum value and the mass loss rate reaches the peak value, respectively.
The red/cyan/magenta lines show the sensitivity curve of LISA/Taiji/Tianqin. The row of purple points at the bottom of the figure denote the contact
frequencies of NS-WD binaries with different WD masses listed in Table~\ref{tabfrequency}. 
Bottom panel: the region $f = 0.01 - 1$ (Hz) and strain $= -20.5 -  -17.5$ enlarged to show the evolution of massive Roche lobe filling NS-WD binaries more clearly. 
}
 \label{fig10}
\end{figure}

In order to investigate the signal-to-noise ratio (SNR) $\rho$
of NS-WD binaries for LISA-type space GW detectors, we calculate the averaged square SNR $\overline{\rho^{2}}$ over the sky location, inclination, and polarization as
\begin{equation}
\overline{\rho^{2}} = \int_{f_{1}}^{f_{2}}\frac{4\cdot \frac{4}{5}fA^{2}(f)}{(P_{\rm n}(f)/R(f))} \rm d (\ln \it f),
\end{equation}
\citep{Moore15,Robson19}, where $f_{1}$ and $f_{2}$ are the lower and upper limits of the integral, respectively.
The factor 4 in the numerator of the integrand comes from the addition of strain noise in the detector arms and the two-way noise in each arm \citep{Larson00}.
We calculate the GW amplitudes $A(f)$ of NS-WD binaries using the phenomenological (PhenomA) waveform model in the Fourier domain \citep{Ajith07, Robson19}.
$A(f)$ is expressed as
\begin{eqnarray}
%
%
A(f) & = & \sqrt{\frac{5}{24}}\frac{G^{5/6}\mathcal{M}^{5/6}}{\pi^{2/3}c^{3/2}R_{\rm b}}f^{-7/6}\,{\rm Hz}^{-1},~~~\it f<f_{\rm m},\\
f_{\rm m} & = & \frac{0.2974\zeta^{2}+0.04481\zeta+0.09556}{\pi (GM/c^{3})}\,{\rm Hz}, \\
\zeta &=& m_{1}m_{2}/M^{2},
\label{ampf}
%
%
\end{eqnarray}
where $\mathcal{M} \equiv m_1^{3/5} m_2^{3/5}/(m_1+m_2)^{1/5}$, and $f_{\rm m}$ is the  GW frequency at the point of merging.
If $f>f_{\rm m}$, the index of the power law relation between $A(f)$ and $f$ changes \citep{Ajith07} and is beyond the scope of this study.
The power spectral density of total detector noise
$P_{\rm n}=\frac{1}{L^{2}}\left[P_{\rm o}+2(1+\cos^{2}(f/f_{\ast}))\frac{P_{\rm a}}{(2\pi f)^{4}}\right]$,
where $f_{\ast}=c/(2\pi L)$, $L=2.5\times10^{9}$ m is the armlength of the detector,
$P_{\rm o}=2.25\times10^{-22} ~\rm m^{2}\left(1+(\frac{2~mHz}{\it f})^{4}\right) ~~ \rm Hz^{-1}$
is the single-link optical metrology noise, and
$P_{\rm a}=9.0\times10^{-30} ~\rm (m~s^{-2})^{2}\left(1+(\frac{0.4~mHz}{\it f})^{2}\right)\left(1+(\frac{\it f}{8~\rm mHz})^{4}\right) ~~Hz^{-1}$
is the single test mass acceleration noise \citep{LISA2018, Robson19}.
$R(f)$ is the transfer function numerically calculated from \citet{Larson00}.
The effective noise power spectral density can be defined as $S_{\rm n}(f)=P_{\rm n}(f)/R(f)$.
For Taiji and Tianqin, we use the sensitivity curve data in \citet{Ruan20} and \citet{Wang19} respectively.

Assuming that the frequency change $\Delta f=f_{2}-f_{1}$ of a binary during one
observation epoch is quite small, the SNR $\overline{\rho^{2}}$ can therefore be written as
\begin{equation}
\begin{split}
\overline{\rho^{2}} = \int_{f_{1}}^{f_{2}}\frac{4\cdot \frac{4}{5}fA^{2}(f)}{(P_{\rm n}(f)/R(f))} \rm d (\ln \it f)\\
\approx \frac{\frac{16}{5}f\cdot \Delta fA^{2}(f)}{f\cdot S_{\rm n}}=\frac{h_{\rm c}^{2}}{h_{\rm n}^{2}},
\end{split}
\end{equation}
where the GW characteristic strain amplitude is defined as
$h_{\rm c}=A(f)\sqrt{\frac{16}{5}f\cdot \Delta f}$, and the detector
noise characteristic amplitude is $h_{\rm n}=\sqrt{f\cdot S_{\rm n}}$.
This definition of the characteristic amplitudes can allow us to calculate the SNR conveniently
from a characteristic strain-frequency diagram.
Figure~\ref{fig10} shows the GW characteristic strain amplitudes of 4U\,1820--30 and other ultracompact X-ray sources with known distances as a function of GW frequency.
For comparison, Fig.~\ref{fig10} also demonstrates the sensitivity curves of the proposed space gravitational
wave detectors LISA, Taiji and Tianqin, as well as other compact binary stars with known orbital parameters, such as detached double WDs and AM CVn stars, including some double
WD X-ray sources.
In order to simplify the model parameters and for convenient comparison, we use the parameters
in Tables~\ref{tab1} and \ref{tab2} to do the calculations.
As seen in Fig.~\ref{fig10}, for an observation of $4$\,yrs, the SNR of 4U\,1820--30 can
be up to $\sim11.0/10.4/2.2$ (for LISA/Taiji/Tianqin respectively).
The SNR of compact binary X-ray sources 4U\,0513-40 and 4U\,1850-087 is
$\sim2.0/1.7/0.33$ (LISA/Taiji/Tianqin) and $\sim1.5/1.2/0.24$ (LISA/Taiji/Tianqin) respectively.
The SNR of other binary X-ray sources is $\lesssim1$.

For NS-WD binaries with larger masses, the changes of  orbital
parameters are quite fast. The SNR can reach a high value in a short observation epoch.
For instance, the SNR of a binary with masses $1.4M_\odot+0.5 {\rm M}_{\odot}$ around contact can reach $\sim27/40/28$ (LISA/Taiji/Tianqin) with one week of observations.
The SNR of an NS-WD binary with masses $1.4 M_\odot+1.0 {\rm M}_{\odot}$ is up to $\sim51/62/114$ in the same observation time.
The evolution track of this binary is shown as a short track in Fig.~\ref{fig10}, where the SNR assumes a 4-year observation.

The longer evolution track of a binary with masses of $1.4 M_\odot+1.25 {\rm M}_{\odot}$ is also shown in Fig.~\ref{fig10},
assuming a shorter observation time of $\sim9$ hours.
Here we assume the distances of these three binaries are $10$\,kpc.
Increasing the distance to  $\sim 100$\,kpc, i.e. the edge of the Galaxy, the detection of GW from these binaries will be marginal.
Combining the changes of GW frequency and amplitude, we may be able to diagnose the evolution stage
and reveal the detailed mass transfer process of contact binaries, if any can be detected by future space GW detectors.

Note that the profile of the NS-WD binary (e.g. $1.4 M_\odot+1.25 {\rm M}_{\odot}$) evolution track
is dominated by the GW frequency evolution (for conservative case, see Eq.\,\ref{eq_gwf}), 
and that the chirpmass of the binary persistently decreases with respect to time.

The accretion parameter $\alpha$ in Eq.\,\ref{eq_adot} is tightly associated with the evolution of chirp mass and GW frequency
and hence significantly affects the GW amplitude.
The influence of
$\alpha$ on the frequency and  strain evolution of a $(1.4 + 1.20) {\rm M}_{\odot}$ NS+WD binary is also
illustrated in Fig.\,\ref{fig10} using purple dashed lines.
This example approximates the critical WD mass for direct collision of NS and WD
in the conservative case $\alpha=1$. $\alpha(t)$ represents time-dependent
WD mass loss and NS mass accretion in our model. We can see that the maximum frequency increases with $\alpha$ as expected
from the evolution of orbital separation (Eq.\,\ref{eq_adot}). The relation between the characteristic strain and GW frequency
is not simply a power-law because the chirp mass is a complicated function of the frequency and its time derivative.

\begin{figure}
\hspace*{-0.35cm}
\centering
\includegraphics[width=9.5cm,clip,angle=0]{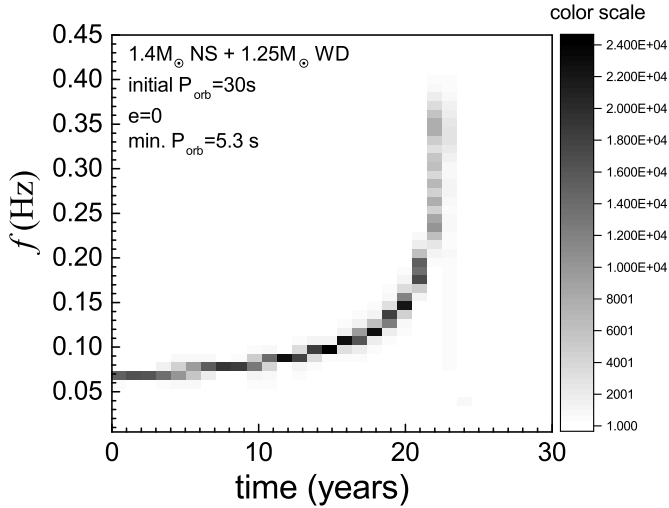}
\caption{The arbitrary power spectrum distribution as a function of GW frequency and evolution time.
}
 \label{fig11}
\end{figure}

Note that Eq.\,\ref{ampf} is valid for the in-spiral evolution of a NS-WD binary purely due to GW radiation.
After onset of Roche lobe overflow, mass transfer will modify the mass-quadrupole evolution, resulting in
a complicated GW amplitude evolution. In order to understand the amplitude evolution, we first calculate the strain $h_{\rm xy}$
for the NS-WD with masses (1.4+1.25)$\rm M_{\odot}$ using the expression for $h_{\rm xy}$ in \citet{Yu15}, then compute the Fourier
transform of $h_{\rm xy}$ to obtain its power spectrum distribution (PSD) over a short time interval.
For this binary, we take the interval $\sim1.0$ yr and time step 0.001 yr.
Finally we obtain the time evolution of the PSD.
In this model, we assume that the binary orbit is already circularized by
the tidal interaction and GWs, so the binary eccentricity  $e=0.0$. Fig.\,\ref{fig11} shows the evolution of the PSD (grey scale) as a function of GW frequency and time.
Note that the color scale is arbitrary.
The PSD evolution would be compared with space-based GW observations to constrain the parameters in our models.

The orbit-spin coupling has some effects on the averaged maximum GW amplitude over one orbital period and
the maximum frequency.
In our computation grids for the coupling (see section \ref{lab_oscoupling}), the maximum GW amplitude
varies by $\sim$27\%, and the maximum frequency varies by $\sim1\%$.
Since we have not yet considered the time-dependent evolution of NS spin and magnetic field, the detailed effect of orbit-spin coupling remains to be addressed. 

The weak GW signal of low-mass UCXBs (e.g., 4U\,1820--30) may be immersed in the confusion foreground of double WDs or other types of signal.
As shown in Fig.~\ref{fig10}, the signals of known double WDs are stronger (AM\,CVn stars are a kind of semi-detached double WD), mainly because they are closer to us (Tables\,\ref{tab1} and \ref{tab2}).
However, the GW frequency for 4U\,1820--30 is $0.00292$\,Hz and is located in the frequency range of  resolvable sources ($f_{\rm GW} \gtrsim 2$\,mHz). This allows us to detect its GW signal by increasing the frequency resolution.
One possible method is to subtract the foreground signals from observation data, which requires a complete understanding of the foreground signal \citep{Webbink98, Nelemans01b, Liu09, Yu10, Ruiter10, Nissanke12, Yu15, Kremer17,Korol18,Lamberts19,Li20,Korol20,Huang20,Breivik20}.

We can estimate the number ($N$) of NS - Roche lobe filling WD binaries from $N=\nu_{\rm m}m_{2}/\dot{m}_{2}$
where $\nu_{\rm m}$ is the binary merger rate.
The NS-WD merger rate for the Galaxy can be approximated as
$\nu_{\rm m}\thickapprox 10^{-5}-10^{-3}$ yr$^{-1}$ as suggested by recent population synthesis computations
\citep{Nelemans01a,Toonen18}, although the rate depends on a wide range of initial conditions and physical processes
(e.g. different common-envelope evolution models and NS natal kick distribution).
For low-mass NS - Roche lobe filling WD binaries like 4U\,1820--30,
we take $m_{2}=0.065~{\rm M}_{\odot}$ and $\dot{m}_{2}=6\times10^{-8}~{\rm M}_{\odot} {\rm yr}^{-1}$, so their number in the Galaxy may
be in the range of $\sim 11-1100$.
For high-mass binaries, if employing $m_{2}=1.25~{\rm M}_{\odot}$ and
$\dot{m}_{2}=8\times10^{-4}~{\rm M}_{\odot}~\rm yr^{-1}$, their number may be in the range of $\sim 0.02-2$.

Our model for the evolution of NS-WD binaries will provide a useful input for binary star population synthesis to
obtain more accurate statistics for these binaries.
We are developing such models to further estimate the Galactic NS-WD merger frequency, their total numbers, and their mass - frequency distributions, and to further calculate their individual and superposed GW signals.

\begin{table}
\caption{The orbital parameters and distances of low-mass ultracompact X-ray binaries.
$P_{\rm orb}$=Orbital period of a binary; $M_{\rm NS}$=Mass of a neutron star;
$M_{\rm WD}$=Mass of a white dwarf at the onset of its Roche lobe; D=Distance.}
\hspace{-0.5mm}
\label{tab1}
\begin{center}
\begin{tabular}{lccccc}
\hline
          Name   & \multicolumn{1}{c}{$P_{\rm orb}$}& \multicolumn{1}{c}{$M_{\rm NS}$} & \multicolumn{1}{c}{$M_{\rm WD}$}& \multicolumn{1}{c}{D}\\
                 & (s)                              & (${\rm M}_{\odot}$)                   & (${\rm M}_{\odot}$)                   & (kpc)             \\
   \hline
4U\,1820--303$^{1,2}$  (NGC6624)      & 685            & 1.0            & 0.0651        & 7.6         \\
4U 0513-40$^{3}$ (NGC1851)       & 1020           & 1.0            & 0.0435        & 12                \\
4U 1543-624$^{4}$                 & 1092           & 1.0            & 0.0405        & ---              \\
4U 1850-087$^{3}$ (NGC6712)      & 1236           & 1.0            & 0.0356        & 8.2               \\
M15 X-2$^{3,5}$ (M15)    & 1356           & 1.0            & 0.0325        & 10.4             \\
XTE J1807-294$^{6}$            & 2406           & 1.0            & 0.0184        & ---                \\
4U 1626-67$^{7}$                 & 2520           & 1.0            & 0.0173        & ---                \\
XTE J1751-305$^{6}$              & 2544           & 1.0            & 0.0171        & ---               \\
XTE J0929-314$^{6}$              & 2616           & 1.0            & 0.0166        & ---             \\
NGC6652B$^{7}$  (NGC6652)         & 2616           & 1.0            & 0.0166        & 9.6                \\
X1832-330$^{8}$  (NGC6652)        & 2628           & 1.0            & 0.0165        & 9.6               \\
4U 1915-05$^{7}$                  & 3000           & 1.0            & 0.0144        & 9               \\
4U 0614+091$^{9}$              & 3060           & 1.0            & 0.0142        & 3.2              \\
\hline
\hline
\end{tabular}
\footnotesize{References: $^1$\citet{Stella87}, $^2$\citet{Guver10}, $^3$\citet{Prodan15}, $^4$\citet{Wang04}, $^5$\citep{Dieball05}, $^6$\citet{Deloye03}, $^7$\citep{intZand07}, $^8$\citet{Deutsch00}, $^9$\citet{Shahbaz08}.
}
\end{center}
\end{table}

\begin{table}
\caption{The orbital parameters and distances of AM CVn stars, detached double white dwarfs (DWDs)
and hot subdwarf (HS) binaries.
$P_{\rm orb}$=Orbital period of a binary; $M_{\rm 1}$=Mass of a primary star;
$M_{\rm 2}$=Mass of a secondary star; D=Distance.}
\label{tab2}
\footnotesize
\begin{center}
\begin{tabular}{cccccc}
\hline
\raisebox{0ex}[0cm][0cm]{Name}& \multicolumn{1}{c}{$P_{\rm orb}$}& \multicolumn{1}{c}{$M_{\rm 1}$} & \multicolumn{1}{c}{$M_{\rm 2}$}& \multicolumn{1}{c}{D}\\
\cline{1-5}
                & (s)            & (${\rm M}_{\odot}$)  & (${\rm M}_{\odot}$)  & (kpc)             \\
   \hline
AM CVn stars                     &                &                 &               &                  \\
\hline
HM Cnc$^{a}$                     & 321.5            & 0.55            & 0.27           & 5                 \\
V407 Vul$^{a}$                   & 569.4            & 0.80            & 0.18          & 1.786              \\
ES Ceti$^{a}$                    & 620.2            & 0.80            & 0.16          & 1.584              \\
SDSS J135154-064309$^{a}$        & 943.8            & 0.80            & 0.10          & 1.317         \\
AM CVn$^{a}$                     & 1028.7           & 0.68            & 0.13          & 0.299            \\
SDSS J190817+394036$^{a}$        & 1085.7           & 0.80            & 0.09          & 1.044        \\
HP Lib$^{a}$                     & 1102.7           & 0.49            & 0.05          & 0.276             \\
PTF1 J191905+481506$^{a}$        & 1347.4           & 0.80            & 0.07          & 1.338              \\
CXOGBS J175108-294037$^{a}$      & 1375.0           & 0.80            & 0.06          & 0.971            \\
CR Boo$^{a}$                     & 1471.3           & 0.67            & 0.04          & 0.337             \\
V803 Cen$^{a}$                   & 1596.4           & 0.78            & 0.06          & 0.347           \\
SDSS J1240-0159$^{1}$            & 2220              & 0.31            & 0.02         & 0.395          \\
GP Com$^{2}$                     & 2794             & 0.5             & 0.02          & 0.075            \\
\hline
Detached DWDs        &                &                 &               &                  \\
\hline
SDSS J065133+284423$^{a}$          & 765.5            & 0.49           & 0.25        & 0.933             \\
SDSS J093507+441107$^{a}$              & 1188.0           & 0.75           & 0.31        & 0.645             \\
SDSS J163031+423306$^{a}$              & 2389.8           & 0.76           & 0.30        & 1.019            \\
SDSS J092346+302805$^{a}$           & 3883.7           & 0.76           & 0.28        & 0.299             \\
\hline
HS Binaries         &                &                 &               &                  \\
\hline
CD-30$^{\circ}$11223$^{a}$          & 4231.8            & 0.79           & 0.54        & 0.337             \\
\hline
\hline
\end{tabular}
\footnotesize{References: $^{a}$ Data is taken from \citet{Kupfer18}
and references therein; $^1$\,\citet{Roelofs05}; $^2$\,\citet{Roelofs07}.
}
\end{center}
\end{table}

\section{Discussion}
\label{sec_discussion}

In the following we consider further our calculations with respect to a) recent work by \citet{Chen20,Tauris18}, b) the physics of mass transfer from white dwarfs, and c) the ultimate fate of contact NS-WD binaries.

We have investigated the orbital evolution of contact NS-WD binaries and their GW radiation.
We find that a LISA-type detector could detect GW from the X-ray source 4U\,1820-30 with a SNR of about 10.
This is consistent with recent results by \citet{Chen20,Tauris18,Nelemans09}.
By numerically solving the structure of white dwarfs, we further find that high mass NS-WD binaries will have distinct evolution tracks in the GW strain-frequency relation which could be resolved by a LISA-type detector.
\citet{Chen20} found that the birthrate of UCXBs is about
(2-2.6)$\times10^{-6}$ $\rm yr^{-1}$, and the number of UCXBs for LISA could be up to 240-320, assuming that all UCXBs evolve from the pre- low-mass X-ray binaries (LMXBs)/intermediate-mass X-ray binaries (IMXBs), to wit, a binary consisting of a neutron star and
a 0.4-3.5 $\rm M_{\odot}$ main sequence star.
The estimated number of  contact NS-WD binaries observable by LISA-type detectors is also a few hundreds, but we use a merger rate of (1-10)$\times10^{-4}$ $\rm yr^{-1}$ for NS-WD binaries to do the estimation.
Detailed population synthesis models are needed to determine the influences of parameters such as star formation
history, initial mass function, and metallicity, physical processes such as common envelope evolution and tidal interaction, as well as formation channels, on the observable numbers of contact NS-WD binaries.

\citet{Tauris18} investigated the evolution of low-mass x-ray binaries and UCXBs and inferred their GW evolution tracks and dynamical masses using the numerical binary stellar evolution tool MESA \citep{Paxton15}.
Their results show the binary nature can be inferred by the strong relation between the GW frequency-dynamical chirp mass, and a precise detection of the chirp provides a new manner to determine the NS mass and constrain its equation of state.
Our results basically agree with \citet{Tauris18} but with a few interesting differences.
They show that mass transfer, or rather mass loss from the white dwarf, is related to the WD equation of state  and to the hydrostatic equation that we have assumed to date.
Tidal interaction and non-conservative mass transfer can also play an important r\^ole in evolution on the GW amplitude -- frequency diagram.
Our results also show that when the initial WD mass is $\gtrsim$ 1.25 $M_{\odot}$, a NS-WD binary may undergo unstable mass transfer and experience runaway coalescence (direct mergers).
The direct mergers will have GW evolution tracks that are quite distinct from those of NS-WD binaries with stable mass transfer.

The non-degenerate envelope of a WD may have a little influence on the orbital evolution of ultra-low-mass NS-WD binaries, but can be neglected in the evolution of high-mass NS-WD binaries.
When the Roche lobe overflow of a NS-WD binary commences, the mean mass density of the overflow matter is about 2.3$\times$10$^{3}$ g cm$^{-3}$ for the binary with masses of (1.0+0.065) ${\rm M}_{\odot}$.
This is comparable to the typical values (e.g. 10$^{3}$ g cm$^{-3}$) for the density at the transition region between the interior degenerate matter of WD and its non-degenerate envelope.
However, for high-mass WDs, e.g. 1.25 ${\rm M}_{\odot}$, the mean mass density of overflow matter can be 1,500 times higher than that of the 0.065 ${\rm M}_{\odot}$ WD.
Depending on the WD temperature, luminosity, opacity, mass and other microphysics, the non-degenerate envelope expands radially by $\lesssim$1\% \citep{Kippenhahn90} at the commencement of mass transfer.
For example, a WD with mass 1.0 ${\rm M}_{\odot}$, radius $8\times10^{-3}$ $R_{\odot}$, and temperature 10$^{7}$ K is extended by $\sim$0.7\%.

As the WD loses more and more mass, it will reach a lower end mass limit at which degenerate pressure no longer dominates the interior pressure to balance the gravity.
The pressure produced by other forces should be involved in the equation of state, e.g. Coulomb interaction or lattice energy.
When a WD mass is lower than this mass limit, the first derivative of radius with respect to mass will become larger than 0 (i.e. $\partial r_{2}/\partial m_{2}>0$).
In this case, the WD will finally become a planet-like object. The mass limit is somewhere between 0.001 and 0.01 ${\rm M}_{\odot}$, depending on chemical compositions \citep{Kippenhahn90}.
Interestingly, \citet{Margalit17} find that the PSR B1257+12 may be such a planetary system, resulting from the merger and tidal disruption of a carbon/oxygen
(C/O) WD - NS binary.
It is not clear whether such a planet-like object rotating around the NS is the finally destiny of such a NS-WD system or whether the planet will also be slowly accreted by the NS companion via Roche-lobe overflow. This deserves further investigation.

\citet{Metzger12} used one-dimensional steady-state models of accretion discs
produced by the tidal disruption of a WD by a NS or a black hole, and find that at the high accretion rates
of $\sim10^{-4}-10^{-1}$ M$_{\odot}$ s$^{-1}$, most of the disc is radiatively inefficient and prone to
outflows powered by viscous dissipation and nuclear burning. For reasonably assumed properties of
disc winds, they show that a significant fraction ($\gtrsim$50-80\%) of the total WD mass is unbound.
\citet{Margalit16} showed similar results by employing time-dependent models of accretion discs with nuclear burning.
We do not here consider a detailed model for the disc evolution. Instead, by using the Eddington limit
as a condition to calculate the time-dependent accretion rate (see section \ref{sec_method}),
we find that for a (1.4+0.1) $\rm M_{\odot}$ NS-WD binary, $\gtrsim$50\% of the lost mass of the WD can be accreted by its
companion NS, while for (1.4+1.25) $\rm M_{\odot}$ NS-WD, the accretion parameter $\alpha$  drops steeply
to $5.4\times10^{-4}$, and then slowly increases as the WD mass loss rate decreases.
The effect of disc mass variation (or the unbound mass) on the GW amplitude may be large if the variation of
mass quadrupole tensor is large, despite the chirpmass of a NS/WD plus disc being small. For instance,
the chirpmass of a 2.0 $\rm M_{\odot}$ NS with a 0.01 $\rm M_{\odot}$ disc is $\sim8.3\times10^{-2}$ $\rm M_{\odot}$.
Future calculations should consider this effect (e.g. \citet{Kocsis11}) but are out of scope here.

Multi-messenger research can be done with the observable explosive transient EM events produced by NS-WD binaries.
Smoothed particle hydrodynamics and hydrodynamical-thermonuclear simulations \citep{Zenati20,Zenati19,Metzger12} showed
that isotropic EM emission from WD-NS mergers may be (subluminous) supernova-like transients.
Depending on model parameters (e.g. wind efficiency, WD mass),
more than $\sim$0.3 $\rm M_{\odot}$  can be ejected with mean velocity $1-5\times10^{4}$ km s$^{-1}$,
nickel ($^{56}$Ni) mass up to $0.006$ $\rm M_{\odot}$,
and  bolometric luminosity up to $\sim6\times10^{40}$ ergs s$^{-1}$ (for a (1.4+0.6) $\rm M_{\odot}$ NS-WD binary).
The peak luminosity may increase to $\sim10^{43}$ ergs s$^{-1}$ due to
shock heating of the ejecta by late-time outflows \citep{Margalit16}.
These results are also supported by \citet{Fernandez19},
who also found a wide outflow velocity range  $10^{2}-10^{4}$ km s$^{-1}$ and a $^{56}$Ni mass up to $\sim$0.01 $\rm M_{\odot}$ for CO/ONe WDs.
Studies of the NS-WD binaries with different masses may also enable us to understand the dependency of
their chemical abundance evolution on the mass transfer process, nuclear burning and magnetic fields.

Subluminous supernova-like transients detected as EM counterparts of GW events may also have detached NS-WD progenitors.
Radio observations indicate that quite a few intermediate mass WD-NS (WD masses $0.1 - 1.1 \rm M_{\odot}$) will undergo a significant mass transfer phase within $10^{10}$ yr as listed in Table\,\ref{tab3}.
We can further study the formation channels of these NS-WD binaries by combining these observational samples with population synthesis.
\begin{table*}
\caption{
Observed neutron star - white dwarf binaries which will merge within $10^{10}$ yr.
$P$ = orbital periods; $e$ = eccentricity;
$d$ = distance; $m_{1}$ = assumed pulsar mass;
$m_{2}$ = estimated median companion mass assuming the inclination angle of $60^{\circ}$;
t$_{\rm m}$ = merging timescale.
The properties of these pulsar+WD binaries can be found in the Australia Telescope National Facility
(ATNF) pulsar catalog (http://www.atnf.csiro.au/research/pulsar/psrcat) \citep{Manchester05}
and references therein.
}
\label{tab3}
\begin{center}
\begin{tabular}{lccccccc}
\hline
         Name (PSR)    &      $P$                 & $e$                    & $d$                  & $m_{1}$         & $m_{2}$        &   $t_{\rm m}$ \\
                       &      (days)              &                        & (kpc)                & ($M_{\odot}$)   & ($M_{\odot}$)  & ($\times10^{9}$ yr) \\
 \hline
         J0348+0432    & 0.10                   & 2.40E-06                 & 2.10                &1.40              & 0.1& 0.83  \\
         J1701-3006B          & 0.14	       & $<$7.00E-05	           & 7.05	         	& 1.35            & 0.14    & 1.5 \\
         J2140-2310A(M30A)  & 0.17	     & $<$1.20E-04	   & 9.20	  	& 1.35     & 0.11    & 3.2 \\
         J1141-6545  	& 0.20	                & 1.70E-01	              & 3.20	            & 1.40            	& 1.00	& 0.62 \\
         J1952+2630	   & 0.39               	& 4.09E-05              	& 10.04	              & 1.35          	& 1.13	& 3.4 \\
         J1748-2446N        & 0.39	     & 4.50E-05	     & 5.5	  	& 1.35     & 0.56       & 6.3   \\
         J1757-5322	  & 0.45                 	& 4.00E-06              	& 1.36	             & 1.40	           & 0.60	& 8.4 \\
\hline
\end{tabular}
\end{center}
\end{table*}

High energy events like $\gamma$-ray bursts from NS-WD binaries may ne rare for two reasons:
1) the prompt and afterglow emission are dimmer because of the relativistic debeaming of off-axis radiation;
2) the accreted mass is small, being a factor of $\sim10-1000$ smaller than the initial WD mass.
The latter could also result in a small number of NS-WD merger-induced black hole formation.
Non-thermal radio transients are also possible from NS-WD mergers.
A typical number of detectable transients is $\sim10$ assuming
ejected mass $\sim 0.3 \rm M_{\odot}$,
mean velocity $\sim 10^{4}$ km s$^{-1}$,
merger rates $\sim10^{-4}$ yr$^{-1}$,
interstellar medium  density  $\sim 1$ cm$^{-3}$,
energy density at 1.4 GHz $\sim 0.1$ mJy,
and the fraction of the energy density behind the
shock imparted to relativistic electrons and magnetic fields are both $\sim 0.1$ \citep{Metzger12}.
Combining GW observations with electromagnetic events, we should be able to determine the EM counterparts of NS-WD mergers and localize them.

The GW signals generated by a NS-WD binary could be interfered by other GW signals.
In the case of an accreting NS, the averaged GW amplitude over one period and sky angle
at NS rotation frequency $f_{\rm s}$ is expressed as \citep{Ushomirsky00}
\begin{equation}
h_{\rm m}=\frac{64}{5}\frac{\pi^{5/2}}{\sqrt{3}}\frac{G}{c^{4}}\frac{Q_{22}f^{2}_{\rm s}}{R_{\rm b}},
\end{equation}
where $Q_{22}$ is the quadrupole moment of a star deformed, for example, by crustal mountains on the NS in LMXBs,
due to thermal asymmetries or magnetic confinement \citep{Haskell15, Bildsten98}.
For neutron stars, the thermal mountain model gives  $Q_{22} \lesssim 10^{40}$ g cm$^{2}$,
and rotation frequency (in LMXBs)  $\sim100-1000$ Hz \citep{Haskell15}.
Assuming that the distance $R_{\rm b}=10$ kpc, we have $h_{\rm m}\lesssim4.5\times10^{-24}$, which is a factor of
$\sim2.4$ smaller than the averaged strain amplitude over one orbit period produced by the contact NS-WD binary
4U 1820-30. The GW frequency of the rotating NS may be higher than 0.043 Hz (period $\sim$23.5 s) determined by
the slowest-spinning radio pulsar so far \citep{Tan18}, which means their GW signals could be observed by
high-frequency bands of LISA-type detectors or sensitive ground-based detectors, e.g. Advanced LIGO and the Einstein Telescope.
If applying this GW emission model to a rotating WD and
unless the quadrupole moments of WDs are much larger than $\sim10^{40}$ g cm$^{2}$, we will get a much smaller GW amplitude $\lesssim10^{-30}$
because WDs have  rotation periods $\gtrsim$300 s \citep{Pshirkov20,Reding20,Kawaler15,Kissin15}.

\section{Conclusion}
\label{sec_conclusion}

In this paper, we investigate the influence of mass transfer on the orbital evolution of contact NS-WD binaries, and the detectability of these binaries using space-based GW detectors, such as LISA/Taiji/Tianqin.
We assume that Roche lobe overflow from the WD will occur after contact occurs, that a fraction of mass is accreted by the NS companion,
and that other mass may be lost from the binary system, depending on how much the mass-loss rate exceeds the Eddington accretion limit.
Our results support the conclusions that mass transfer will significantly affect the orbital evolution,
and that the majority of the NS-WD binaries will avoid direct coalescence because of the strong resistance
of mass transfer-induced orbital expansion to  GW-induced orbital decay.
Instead, they evolve to a minimum orbital separation
and then recede from each other.
This study also provides input for binary population synthesis of compact binaries in order to investigate their progenitor populations, current numbers and merger rates.

Our main conclusions are as follows.
\begin{enumerate}
\item The contact GW frequency, corresponding to two times  the reciprocal binary orbital period, at which the WD companion of the binary just fills
its Roche lobe, is in the range  $(0.0022637 - 0.7194)$\,Hz for WD masses  $(0.05 - 1.4) {\rm M}_{\odot}$.

\item After contact, WD companions with masses less than $\sim 1.25  {\rm M}_{\odot}$ (assuming $m_{\rm NS}=1.4 {\rm M}_{\odot}$) will
go through a stable mass transfer (mass loss), and evolve towards a maximum GW frequency.
After that, the WD mass loss rate of will decrease to approach the rate calculated in Eq.~\eqref{eq_balancemdot}.
When the masses of WDs are greater than $\sim1.28 {\rm M}_{\odot}$, their mass loss rates are certainly unstable, and the binaries will most likely coalesce directly.
For WDs in a narrow mass range $\sim (1.25 - 1.28) {\rm M}_{\odot}$, NSs will come into direct contact with their WD companions, although the WD mass loss is stable.
The WD mass range becomes $m_{\rm WD}\ \sim (1.23-1.24) {\rm M}_{\odot}$ when changing the NS mass to 3.0 $\rm M_{\odot}$.
Tidal deformation has a negligible influence on the mass boundary.

\item Our model can simulate the orbital evolution of the ultracompact binary X-ray source 4U\,1820--30 and other low-mass binary X-ray sources.
In terms of GW radiation, each NS-WD pair traces a unique evolutionary path in strain - frequency space.

\item With the current design of space-based GW detectors LISA/Taiji/Tianqin, the SNR of 4U\,1820--30 ($(1.0+0.065) {\rm M}_{\odot}$ NS-WD binary) could be up to $\sim 11.0/10.4/2.2$ for a 4-year observation.
For NS-WD binaries with larger masses, the changes of their orbital parameters are quite fast.
The SNR can reach a high value in a short observation epoch.
For instance, assuming a distance $10$\,kpc, the SNR of a binary with masses $(1.4+0.5) {\rm M}_{\odot}$ could reach $\sim27/40/28$ in an observation time of one week.
The SNR of a binary with masses $(1.4+1.0){\rm M}_{\odot}$ could reach $\sim51/62/114$ in the same observation time.
For a 4-year observation, the $(1.4+1.0){\rm M}_{\odot}$ binary could be observed to evolve in the GW  strain-frequency plane.
Over the same 4-year interval, a $(1.4+1.25){\rm M}_{\odot}$ mass binary would show a longer strain-frequency evolution track.

Combining the measurement of the changes of GW frequency and amplitude,
we will be able to determine the evolution stage of neutron star white dwarf binaries.
\end{enumerate}

\section*{Acknowledgments}
We thank the referee, Hagai Perets, for helpful and constructive comments. 
This work was supported by the National Natural Science Foundation of China
(Grant Nos 11673031, 11690024), the Strategic Priority
Research Program of the Chinese Academy of Sciences (Grant No. XDB23040100), and
the National Key Research and Development Program of China (No. 2020YFC2201400).
This work was also supported by the Open Project Programme of the key Laboratory of Radio Astronomy,
CAS and Cultivation Project for FAST Scientific Payoff and Research Achievement of CAMS-CAS.
Research at the Armagh Observatory and Planetarium is grant-aided by the N. Ireland
Department for Communities.

\section*{DATA AVAILABILITY}
Some or all data, models, or code generated or used during the study
are available from the corresponding author by request.

\bibliographystyle{mnras}
\bibliography{gwrlof}

\label{lastpage}

\end{document}